\documentclass[usegraphicx,useAMS,usenatbib]{mn2e}
\usepackage{times}
\usepackage{subfigure}

\title[Detecting variability in Time-Series Data]
{Detecting Variability in Massive Astronomical Time-Series Data I:
application of an infinite Gaussian mixture model
}
\author[M.-S. Shin et al.]
  {Min-Su~Shin,$^1$\thanks{E-mail: 
  msshin@astro.princeton.edu,
  sekora@math.princeton.edu,
  byun@yonsei.ac.kr}
  Michael~Sekora$^2$
  and Yong-Ik~Byun$^3$\\
  $^1$Princeton University Observatory, Peyton Hall, Princeton, NJ 08544-1001, USA\\
  $^2$Program in Applied and Computational Mathematics, Princeton University, Princeton, 
  NJ 08540, USA\\
  $^3$Department of Astronomy, Yonsei University, Seoul, 120-749, Korea
  }
\date{Released 2008 Xxxxx XX}

\newcommand{\Ratio}{$\sigma/\mu$ } 
\newcommand{\Con}{{\it Con }} 
\newcommand{\Neu}{{$\eta$} } 
\newcommand{\J}{{\it J }} 
\newcommand{\K}{{\it K }} 
\newcommand{\AoVM}{{\it AoVM }} 

\begin{document}

\date{Accepted ... Received ..; in original form ..}

\maketitle

\begin{abstract}
We present a new framework to detect various types of variable objects within massive astronomical 
time-series data. Assuming that the dominant population of objects is non-variable, 
we find outliers from this population by using a non-parametric Bayesian clustering algorithm 
based on an infinite Gaussian Mixture Model (GMM) and the Dirichlet Process. 
The algorithm extracts information from a 
given dataset, which is described by six variability indices. 
The GMM uses those variability indices to recover clusters that are described by six-dimensional 
multivariate Gaussian distributions, 
allowing our approach to consider the sampling pattern of time-series data, 
systematic biases, the number of data points for each light curve, and photometric quality. 
Using the Northern Sky Variability Survey data, 
we test our approach and prove that the infinite GMM is useful at detecting variable objects, 
while providing statistical inference estimation that suppresses false detection. 
The proposed approach will be effective in the exploration of 
future surveys such as GAIA, Pan-Starrs, and LSST, which will produce massive time-series 
data.
\end{abstract}

\begin{keywords}
 stars\ -- variables: other\ -- methods: data analysis, statistical
\end{keywords}

\section{Introduction}

Time-domain astronomy has resulted in a variety of discoveries such as 
gamma-ray bursts and supernovae. 
These kinds of transient phenomenon have made it possible to understand a rare 
stage 
of stellar evolution. Moreover, variable stars have been key objects for 
investigating stellar populations, the structure of the Milky Way, and 
the expansion of the universe \citep{bono05}.

Despite its long history and contribution to astronomy, the study of variable sources is 
not complete yet. 
As \citet{paczynski00} emphasised, there might be unknown 
variable sources. Moreover, known variable objects are not well understood 
\citep{eyer07}. Recently, several surveys revealed a large number of 
variable sources as 
byproducts \citep{paczynski01}. 
Even more new variable sources are expected to be 
discovered in future surveys \citep[e.g.][]{walker03}.

A common approach in the study of variable sources consists of detection, 
analysis in the time domain, analysis in the phase domain with period estimation, and 
classification \citep{eyer05,eyer06}. For each step, various methods have been proposed 
and tested in several projects. One example is a set of variable stars from the 
MACHO project \citep{cook95} where variable objects are 
selected by using chi-square statistics, and the periods of these objects are derived from 
the method explained by \citet{reimann94}. The MACHO project also uses 
a power spectrum of the time-series data to separate out a specific 
kind of a variable star from others \citep{alcock95}. In addition, 
RR Lyrae have been investigated with their distinctive colour and absolute 
magnitude \citep{alcock96}, or visual inspection of light curves 
\citep{alcock97} in the MACHO project.

Period estimation and classification of variable sources have been intensively 
examined by various methods. 
Period determination has been 
tested for data with diverse types of light curves 
\citep[e.g.][]{reimann94,akerlof94,czerny98,shin04}. 
Classification has been explored by using 
statistical tools, including machine learning algorithms 
\citep[e.g.][]{eyer02,belokurov03,belokurov04,debosscher07,
willemsen07,mahabal08}.

However, the general method of variability detection has not been well 
investigated, and a typical method is usually based on a simple probability 
test that is optimised for specific variability types or data \citep[e.g.][]{sumi05}. 
Detection algorithms cannot be separated from the factors that determine 
sampled time-series data: variability types, observation cadence, quality cuts of 
data samples, noise patterns, systematic biases, etc. 
Detecting any type of variable 
object depends on the data we have and how we measure variability. 
Therefore, variability detection has to be a data-oriented process
without dependence on assumptions about the given data.

General variability detection methods must be based on the following requirements 
\citep[see][for a discussion]{eyer06}. First, the method has to 
recover a broad range of variability types. Particularly, the detection method 
needs to be able to recover a new type of variability. 
Second, a probabilistic inference has to be derived in order to help people 
estimate detection reliability. As the amount of data increases, controlling 
the detection of a false positive  becomes important. Third, it is critical 
for the detection method to deal with a variety of data sets such as the number of 
data points, uncertainties in the measured data, and time-sampling patterns \citep{carbonell92}. 
Even in a single survey project where one defined cadence is valid, people 
can adopt different values for data quality cuts 
because of varying observing environments and different properties of 
each observation field such as precision of photometry. Such differences can 
result in a heterogeneous distribution of data points.

In this paper, we propose a new framework to detect a broad range of variability 
within massive time-series data. We employ an unsupervised Bayesian machine learning 
algorithm which 
uses an infinite Gaussian Mixture Model (GMM) with the Dirichlet Process (DP) 
\citep[see][for an example of the GMM in astronomy]{kelly04,debosscher07,bamford08}. 
In this context, separating variable objects from non-variable ones can be regarded as 
a clustering problem \citep{jain99}, or detecting outliers from the cluster of 
non-variable objects \citep{cateni08}.

We adopt six variability indices that are measured from light curves in the time domain 
and used as input features for clustering with the infinite GMM. These indices 
summarise the systematic structure of an individual light curve in the time domain. 
All of the variability indices are estimated by considering the photometric 
uncertainty and number of data points in each light curve. 
Because these indices cover different features of data which are associated with 
variability types, sampling patterns, etc., the GMM 
encompasses a broad range of variability types. 
Using a combination of multiple indices has been suggested by \citet{shin07}. 

In our approach, the infinite number of components\footnote{
We use {\it component} as the same term as {\it cluster} and {\it group} in this paper.} 
which are described by multivariate Gaussian 
distributions represent the six-dimensional space spanned by the variability indices. 
Unlike the GMM used in other astronomical research, our method is based on 
the DP which makes our approach non-parametric by constructing the prior probability from 
the given data\footnote{Even in a 
{\it non-parametric} Bayesian method, a parametrised model is still used as in a 
{\it parametric} Bayesian method. The difference is that 
the {\it parametric} method has a fixed number of parameters so that 
the complexity of the model is fixed. Meanwhile, the number of parameters changes 
according to the complexity of the given data 
in the {\it non-parametric} method \citep{walker99,muller03,jordan05}.}
\citep[see][for an application of the DP in astronomy]{chattopadhyay07}.
The clusters of data points are self-recognised by Bayesian reasoning and 
the DP. Like other unsupervised learning methods, this method fully 
exploits all of the information in the data. 

After 
the infinite GMM is found for the given data, statistical inference measures 
how convincingly candidates of variable objects can be separated out, which 
helps one quantify the reliability of recognising variable sources. 
The only assumption made by this approach is that the largest cluster of the GMM 
is a cluster of non-variable sources. 
Therefore, the GMM works well when data has a 
dominant cluster of non-variable objects as we generally find in astronomical time-series data.

In this paper, we show how to use the infinite GMM with 
the DP for variability detection. 
Using six variability indices of 
time-series data from 
the Northern Sky Variability Survey (NSVS) \citep{wozniak04}, 
we find the largest cluster that should represent a 
cluster of non-variable sources. 
The reliability of the non-variable cluster is tested for the size and properties 
of the data. We use the identified clusters to separate out variable source candidates 
from the data.

The paper is organised as follows. In \S2, we explain the NSVS data, 
variability indices, and infinite 
GMM with the DP. 
The application of the GMM is given for the sample data in \S3, showing 
the reliability and stability 
of the found non-variable cluster that is examined for the size and properties of the data. We 
explain how to measure the significance of variability in \S4. 
The discussion and conclusions are given in the last section. In the appendix, we present 
the basic mathematical explanation of the infinite GMM with the DP.

\section{Method}

\subsection{Test Data}

\begin{table*}
\begin{minipage}{126mm}
\caption{NSVS Fields of the set A.}
\label{tab:fields}
\begin{tabular}{@{}crrccc}
\hline
Name & Galactic $l$ & Galactic $b$ & Number of frames & 
Number of objects & Limiting photometric scatter \\
\hline
065d & 78.0 & -8.0 & 235 & 55051 (46925) & 0.030 \\
087a & 49.0 & 10.0 & 299 & 54749 (47510) & 0.029 \\ 
088d & 60.0 & -8.0 & 196 & 55465 (48155) & 0.030 \\
135b & 16.0 & -6.0 & 106 & 55399 (41142) & 0.043 \\
135d & 27.0 & -9.0 & 102 & 55039 (43480) & 0.034 \\
\hline
\end{tabular}

We present the numbers of objects that have more than 15 good data points in 
the parenthesis.
\end{minipage}
\end{table*}

\begin{figure*}
\includegraphics[scale=0.55]{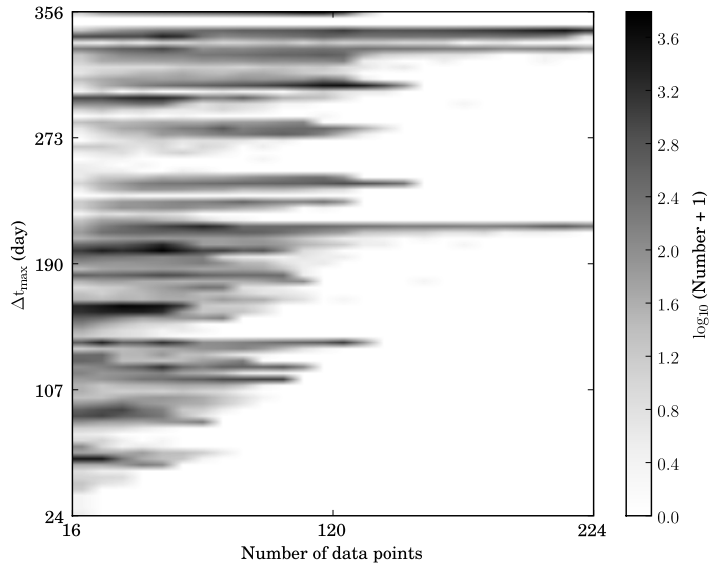}
\caption{The number of data points and maximum time span for set A. 
Five test fields show a broad distribution in the number of data points. 
The maximum time span also reveals a broad distribution 
that does not depend on the number of data points.}
\label{fig:data}
\end{figure*}

We use light curves that have more than 15 good photometric data points 
in the NSVS database\footnote{http://skydot.lanl.gov/nsvs/nsvs.php}. 
A systematic search of the various kinds of variable sources 
has not been carried out with the NSVS data. But photometric quality control is well understood, 
and we can use the large amount of photometric data that 
allows us to recover new variable sources. 
We select five NSVS fields (065d, 087a, 088d, 135b, 135d) 
that have the largest number of objects in those fields. 
The basic information for those fields is given in Table \ref{tab:fields} \citep{wozniak04}. 
We call these data set A. 
As \citet{wozniak04} suggested in their Table 3, we use only good photometric data points, 
which avoid any artifacts from observation and data processing, for each object. 
This photometric quality cut prevents any effects from spurious data points. When 
we limit samples that have more than 15 good photometric points, the number of objects 
from 
each field is about 45,000. The total number of objects is 227,212.

As shown in Figure \ref{fig:data}, 
the number of data points and time-scale of light curves has a 
broad range. The number of data points in the light curves 
affects the uncertainties in the variability indices that will be 
explained in the following section. 
Furthermore, the number of data points is decided by a sampling pattern for each field 
as well as the selection of good photometric data. 
The extractable information is also subject to the time scale of 
the light curve. In the case of periodic variability, the Nyquist frequency is 
an important 
measurement \citep{koen06}. However, if we consider a general type of variability and irregular 
sampling, it is useful to 
examine the distribution of the maximum 
time span among data points. 
Only a tiny fraction of the light curves covers a 
time span of about 300 days with more than 
100 data points, where a dominant fraction of the 
data has less than 60 data points. 

\begin{table*}
\begin{minipage}{126mm}
\caption{NSVS fields of the set B.}
\label{tab:fields_test}
\begin{tabular}{crrccc}
\hline
Name & Galactic $l$ & Galactic $b$ & Number of frames & 
Number of objects & Limiting photometric scatter \\
\hline
045a & 99.0 & -6.0 & 304 & 54455 (45551) & 0.032 \\
064a & 66.0 & 9.0 & 308 & 54320 (46392) & 0.028 \\
089a & 64.0 & -13.0 & 289 & 54363 (46639) & 0.025 \\
112a & 43.0 & -9.0 & 228 & 54334 (43191) & 0.031 \\
135a & 23.0 & -2.0 & 112 & 54096 (40601) & 0.037 \\
157d & 10.0 & -10.0 & 61 & 54391 (4838) & 0.031 \\
\hline
\end{tabular}

The numbers in the parenthesis represent objects that have more than 
15 good data points. We use a part of the data from field 157d.
\end{minipage}
\end{table*}

\begin{figure*}
\includegraphics[scale=0.55]{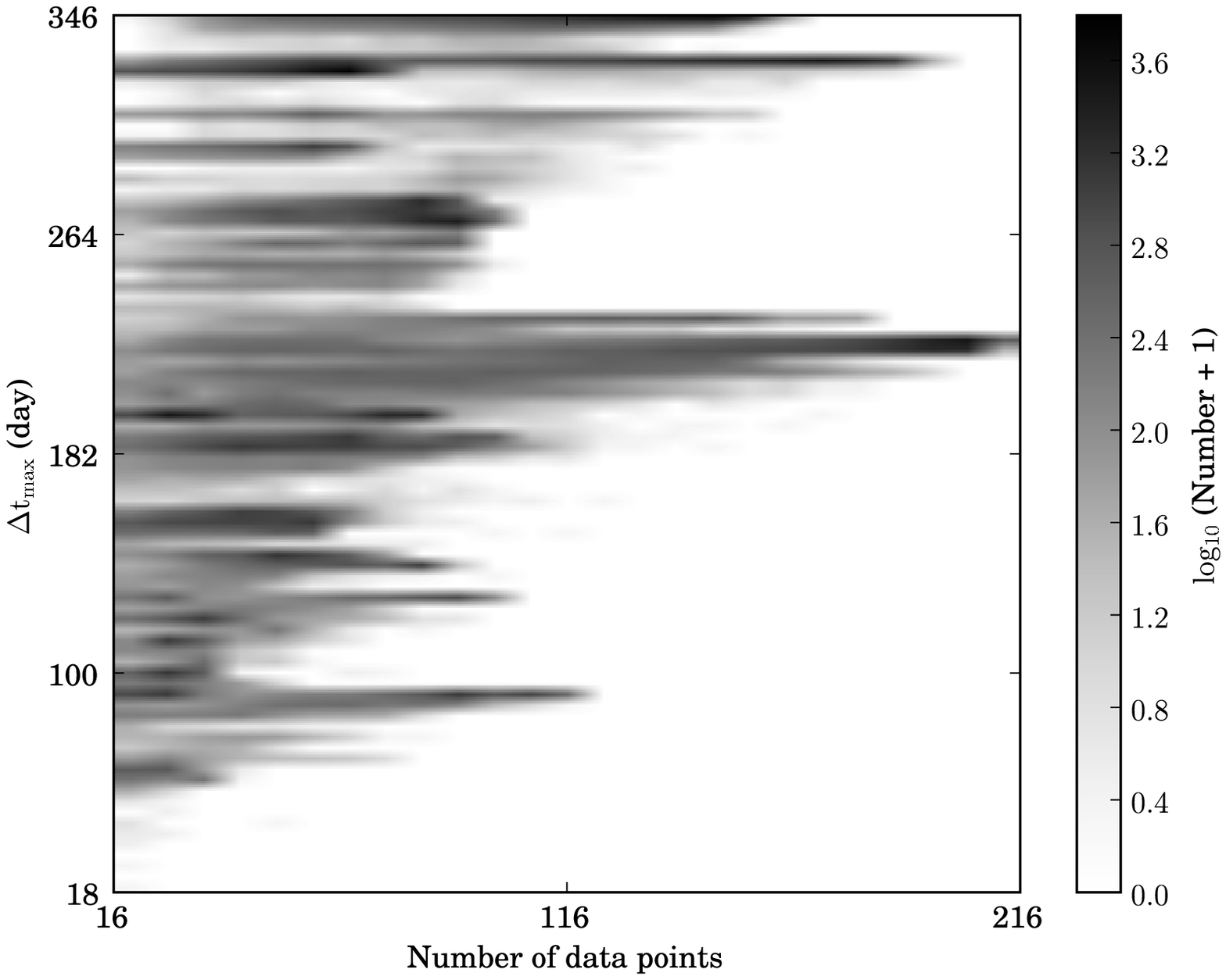}
\caption{The number of data points and maximum time span for set B. 
Compared with set A, the light curves of set B have a smaller number of data 
points. Set B also covers a smaller time span than set A.}
\label{fig:data_test}
\end{figure*}

We also extract set B which has 
the same number of light curves from 
six different fields of the NSVS as set A. 
The differences between these two sets arise from the overall number of 
frames being larger in set B than in set A (Table \ref{tab:fields_test}). 
However, using only good photometric data points as we do with set A \citep[see][Table 3]{wozniak04} 
makes 
the light curves of set B includes less data points 
than set A. Additionally, set B has fewer light curves with a large time-span and 
many data points as shown in Figure \ref{fig:data_test}.

\subsection{Variability indices}

\begin{figure*}
\includegraphics[scale=0.4]{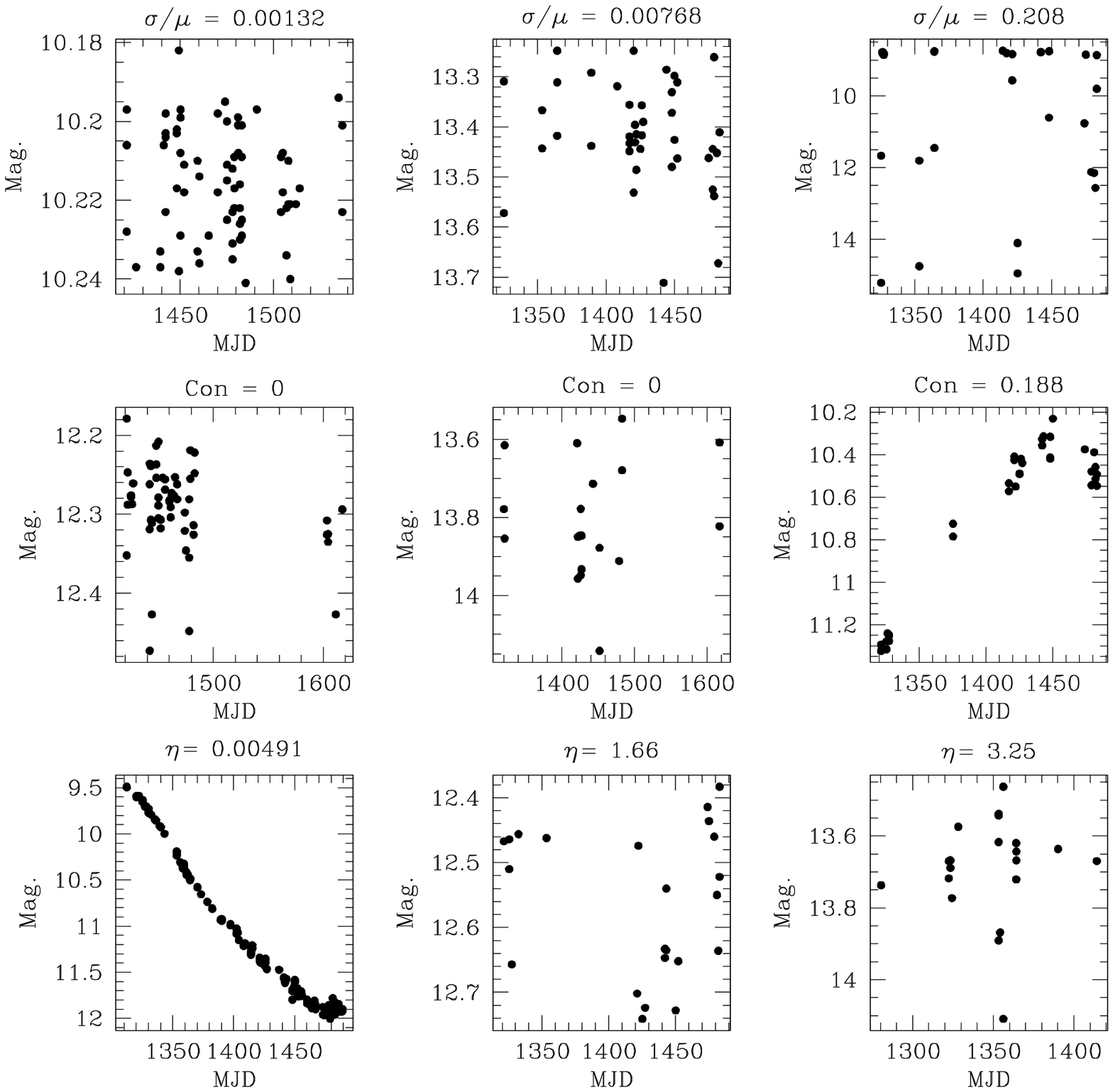}
\includegraphics[scale=0.4]{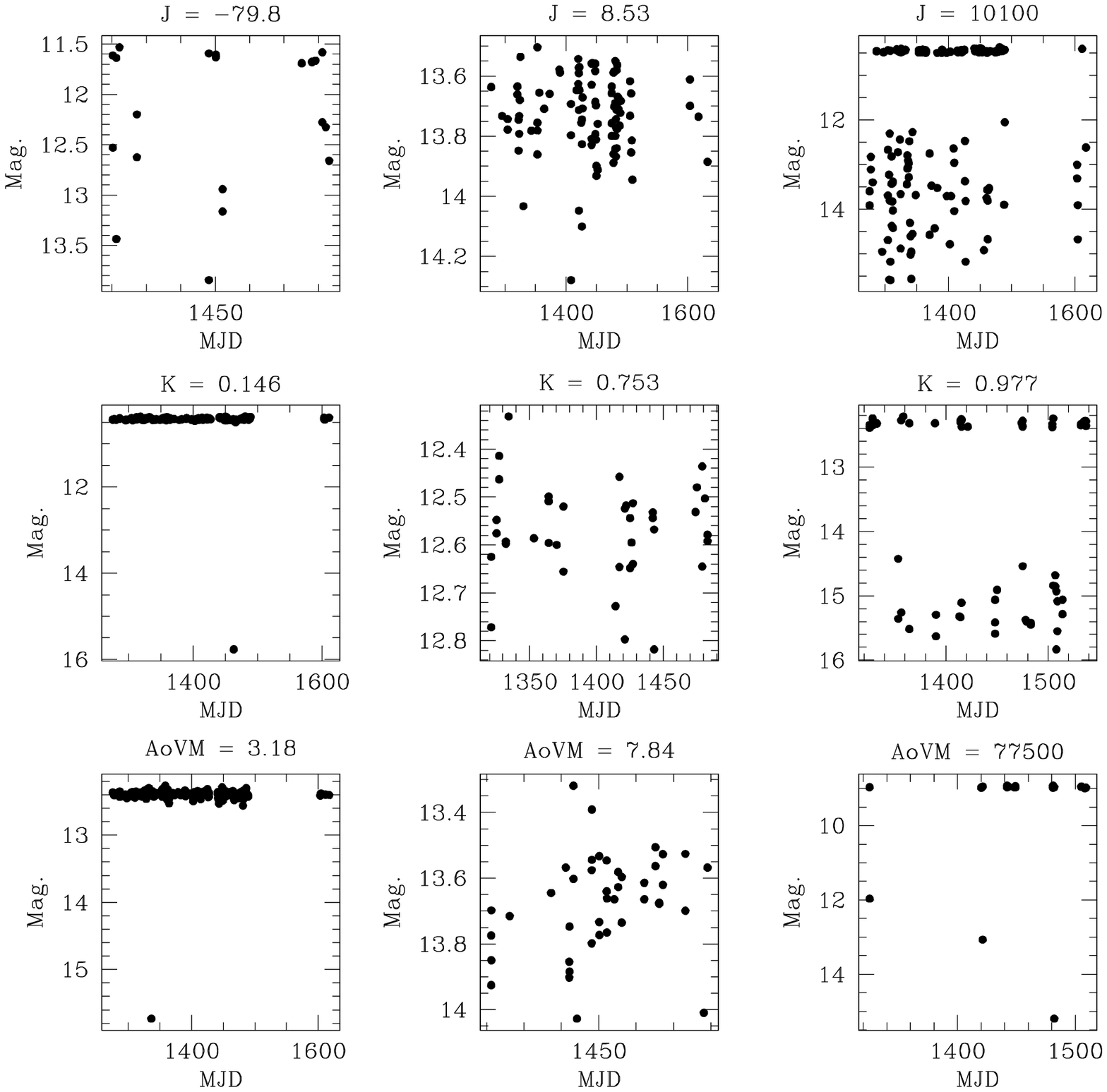}
\caption{Light curves with minimum, median, and maximum parameter values. 
The left three columns 
present light curves with minimum, median, maximum values of 
$\sigma/\mu$, $Con$, and $\eta$. 
The right three columns are the same light curves but with $J$, $K$, and $AoVM$. 
Even among these examples, we recognise a light curve of a variable star with 
the smallest $\eta$, which is a known variable star BS Her \citep{nassau61}, because 
of its monotonic increasing magnitude. 
The light curve with the largest value of $Con$ 
shows a systematic variation that may be a variable star which 
corresponds to the infrared source IRAS 18402-1742 \citep{helou88}.}
\label{fig:lc_example}
\end{figure*}

\begin{figure*}
\includegraphics[scale=1.0]{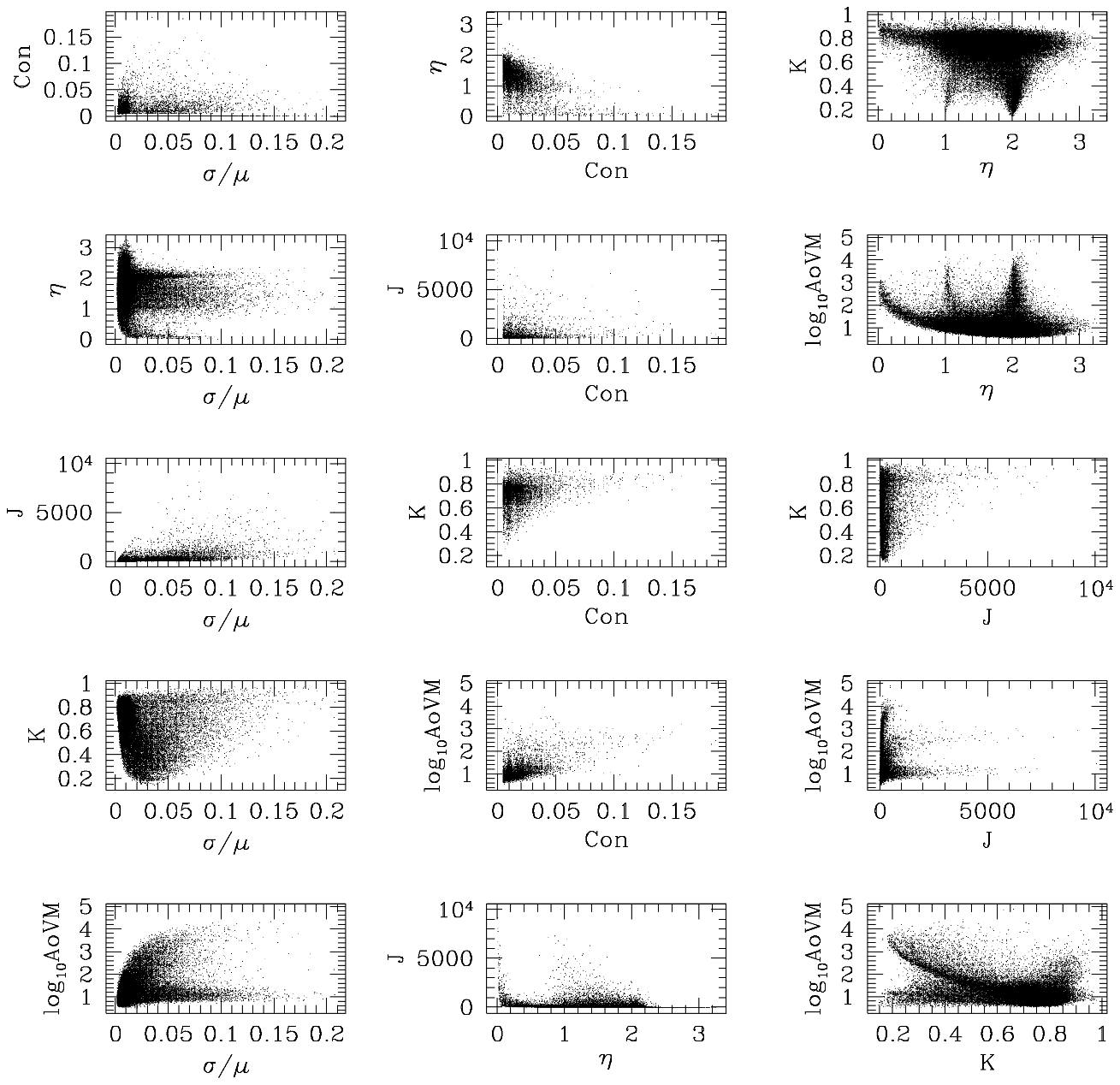}
\caption{Distribution of variability indices for the set A. Each variability 
index describes a different feature of light curves in time domain. But we find the 
existence of the strong concentration 
of data in this six dimensions of the variability indices. 
It implies that the GMM can definitely 
find a dominant cluster of non-variable sources, 
while also separating outliers from the dominant cluster as separate minor clusters.}
\label{fig:var_time}
\end{figure*}

Below we define six variability indices 
($\sigma/\mu$, $Con$, $\eta$, $J$, $K$, $AoVM$) that are obtained from 
light curves in the time domain. The simplest index of variability is 
the ratio of the standard deviation to the sample mean magnitude
\begin{equation}
\frac{\sigma}{\mu} ~=~ \frac{\sqrt{ \sum_{n=1}^{N}(x_{n} ~-~ \mu)^2 / (N ~-~ 1) }}
{\sum_{n=1}^{N}x_{n} / N},
\end{equation}
where $n$ is an index over the relevant data points and $N$ is the total 
number of data points in each light curve. 
When this ratio is large, the light curve may have strong variability. We 
note that this ratio is not correspondent to a flux ratio because 
magnitude is a logarithmic unit. 

However, \Ratio does not describe 
detailed features of variability. Therefore, we find three consecutive 
points that are at least 2 $\sigma$ fainter or brighter than the median magnitude 
in order to trace a continuous variation in the data points. The number 
of consecutive series is normalised by $(N - 2)$, and is called 
\Con. This measurement was used in \citet{wozniak00}. 

The systematic structure of the light curves is also quantised by the ratio of the mean 
square successive difference to the sample variance \Neu \citep{vonneumann41}:
\begin{equation}
\eta ~=~ \frac{\delta^{2}}{\sigma^{2}} ~=~  
\frac{\sum_{n=1}^{N-1}(x_{n+1} ~-~ x_{n})^2 / (N ~-~ 1)}{\sigma^{2}}. 
\end{equation} 
This von Neumann ratio was suggested to test the independence of random variables in 
successive observations particularly on a stationary Gaussian distribution. 
When there is a strong positive (negative) serial correlation between sequential data points, 
this ratio is small (large). In short, if serial correlation exists, the ratio is 
significantly high or small \citep{panik05}. 
The distribution of \Neu has been extensively investigated for a stationary Gaussian distribution 
\citep[e.g.][]{bingham81}, and its sample average and variance are well known 
\citep{williams41}. But the properties of $\eta$ are 
not simple for astronomical time-series data because they are irregularly sampled and 
do not follow a simple known distribution such as a stationary Gaussian distribution. 

Three additional indices are adopted from concepts that have been developed in 
astronomy community. 
\J and \K are suggested by \citet{stetson96}. We use the following definition 
that uses only a single photometric band:
\begin{equation}
J ~=~ {\sum_{n=1}^{N-1}sign(\delta_{n}\delta_{n+1})\sqrt{\vert\delta_{n}\delta_{n+1}\vert}},
\end{equation}
\begin{equation}
K ~=~ {\frac{{1/N}\sum_{n=1}^{N}\vert\delta_{n}\vert}{\sqrt{1/N \sum_{n=1}^{N}\delta_{n}^{2}}}},
\end{equation}
where $\delta_{n} ~=~ \sqrt{N/(N-1)} (x_{n} - \mu)/e_{n}$ 
which has a photometric error for each 
data point $e_{n}$ and $sign(\delta_{n}\delta_{n+1})$ is the sign of 
$\delta_{n}\delta_{n+1}$. Finally, we measure the analysis of variance 
(ANOVA) statistic which is useful for discovering periodic signals \citep{czerny96}. 
The maximum value of the ANOVA represented by \AoVM is 
used to measure the strength of periodicity. 
Even though the corresponding period can be incorrect, 
the \AoVM is still a valuable quantity that infers periodicity 
\citep{shin07}.

Figure \ref{fig:lc_example} shows light curves with 
variability indices that have the minimum, median, and maximum values 
across all of the 227,212 light curves in set A. None of the light curves 
occur more than once in Figure \ref{fig:lc_example}. These examples 
prove that different 
variability indices catch different features of light curves. The light curve 
of the infrared source IRAS 18402-1742 \citep{helou88} has the largest value of $Con$. We 
suspect that the variation of the light curve is real, and the star might be a long-period 
variable star. The light curve with the minimum value of $\eta$ 
corresponds to a known variable star 
BS Her \citep{nassau61}. As we expect, the positive serial correlation in magnitude 
has a small $\eta$ in this light curve.

The variability indices complement each other by picking up different 
features of the light curves. As shown in Figure \ref{fig:var_time}, 
even though we notice some structure in the distributions for each 
two-dimensional projection of the original six-dimensional 
space, the indices do not have 
a strong correlation with each other. If a dominant fraction of 
light curves is simply from a Normal distribution, we would 
see only one simple structure in all plots. Since 
light curves of non-variable objects are not random samples from a Normal distribution, 
each plot shows more complicated structures which imply the existence of variable objects. 
Any structures will be defined as a separate cluster by the GMM. 
However, the strong concentration of data in each plot implies 
the existence of a dominant cluster 
of non-variable objects. Additionally, combining 
multiple indices helps us suppress the false detection of variable sources 
while not missing any possible features in the variability.

\subsection{GMM}

In the infinite GMM based on the DP, the distribution of 
mixture component members is described by a multivariate Gaussian 
distribution while the distribution of all objects is described 
by a mixture of Gaussian distributions defined by the stochastic 
DP. Each of the $M$ component distributions has 
the following form:
\begin{equation}
p_{m}(x) = \frac{1}{(2\pi)^{\gamma/2}\vert\Sigma_{m}\vert^{1/2}} 
\exp(-\frac{1}{2}({\bf x} - {\bf \mu_{m}})^{T}\Sigma_{m}^{-1}({\bf x} - {\bf \mu_{m}})),
\label{eq:gaussian}
\end{equation}
where $m$ is an index over $M$, ${\bf x} = (\sigma/\mu, Con, \eta, J, K, AoVM)$ 
is a 6-dim vector of parameters, and $\gamma$ is the number of parameters 
(in our case $\gamma=6$). Furthermore, ${\bf \mu_{m}}$ is a $6$-dim vector of mean 
values (i.e., mixture centres), and $\Sigma_{m}$ is the covariance matrix of 
the Gaussian distribution associated with the $m$th mixture component. 
The problem is how to find a weighting for each mixture component $w_{m}$ and 
its respective $\mu_{m}$ and $\Sigma_{m}$ such that the final distribution 
of all objects is given by:
\begin{equation}
p({\bf x}) = \sum_{m=1}^{M} p_{m}({\bf x}) w_{m}.
\end{equation}
The DP is used to estimate $w_{m}$, $\mu_{m}$, and $\Sigma_{m}$ 
and is explained in Appendix A.

When loading a given data set, one also specifies initial values for 
hyper-parameters used in the clustering algorithm. These hyper-parameters 
include the number of iterations to be taken by the algorithm to ensure 
convergence to a stable model, number of initial mixture components $M$ 
to which data is assigned, and concentration $\alpha$ which can be 
thought of as the inverse variance of the DP. In all of the 
models presented in this paper, the number of iterations is 100 even 
though convergence can be seen in as little as 10 iterations, $M = 60$ 
initially, and $\alpha = 1$. Convergence is defined by $M$ reaching some 
consistent value despite the algorithm continuing to iterate. To ensure that 
the clustering algorithm identifies all relevant features (i.e., number of 
mixture components), it is possible to initialise the algorithm with $M = N$, 
(i.e. the number of data points) 
which is the highest possible complexity. Since we are mainly concerned 
with identifying one central cluster (i.e., non-variable objects), this 
computationally expensive procedure is unwarranted. With the data set 
and hyper-parameters loaded, the algorithm first creates an empty Gaussian 
distribution $G_{0}$ of mixture components $M$ with a conjugate Gaussian-Wishart 
prior such that the mean vector is drawn from a Gaussian distribution and 
precision matrix (i.e., the inverse covariance matrix $\Sigma_{m}^{-1}$) 
is drawn from a Wishart distribution. Second, the algorithm randomly initialises 
the mixture component assignments $z = \lceil R_{N} M \rceil$, where $R_{N}$ is 
a $N$-dim vector with entries that are uniformly distributed. 
By using $\alpha$, $G_{0}$, and $z$, data points ${\bf x}$ are added to the 
mixture components. As the algorithm iterates to convergence, this initial 
assignment matters little because conditional probabilities are computed for 
each data point with respect to each of the $M$ active mixture components. 
Lastly, a collapsed Gibbs sampler runs for the specified number of iterations 
while also iterating over $N$. We implement this algorithm by using 
MATLAB\footnote{MATLAB is a registered trademark of The Mathworks.}.

\section{The largest cluster as non-variable objects}

Since the largest cluster must represent non-variable light curves, and 
our GMM with the DP is a data-driven unsupervised 
machine learning algorithm, the properties of the largest Gaussian mixture must be dependent 
on the input data. Therefore, we examine the dependence of results on the size and 
properties of the input data. 

\begin{figure}
\includegraphics[scale=0.4]{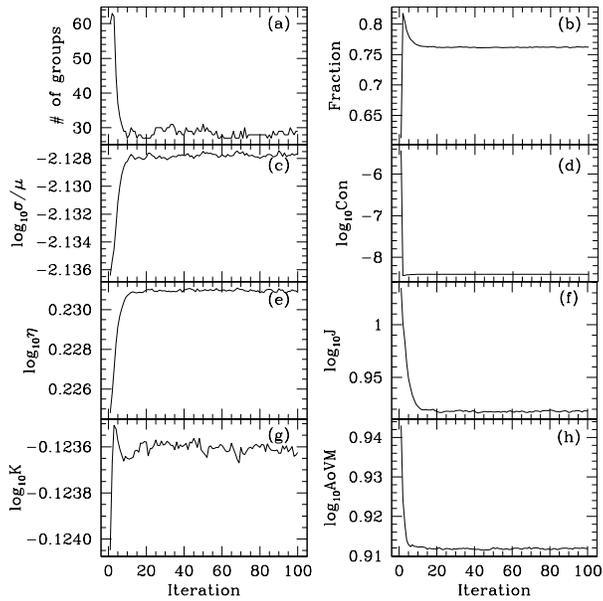}
\caption{The change in cluster properties for set A as a 
function of the number of iterations. (a) The total number 
of identified clusters converges to 29 after 10 iterations. (b) The number 
of data in the largest cluster reaches 82\%, but converges to 
76\% after 100 iterations. (c - h) The center of the largest 
cluster does not show a significant change after the 10th 
iteration. It means that the small changes in the number of clusters after 
the first 10 iterations does not affect the Gaussian model of the 
largest cluster.}
\label{fig:gmm_time_history}
\end{figure}

\subsection{Results of set A}

The GMM of set A is composed of 29 mixture components 
where one cluster dominates the data. 
As shown in 
Figure \ref{fig:gmm_time_history}, the number of mixture components quickly 
converges to about 29 after 10 iterations. Moreover, the centre of 
the dominant cluster remains stationary. The 
largest cluster is populated by 76.2\% of the input data, and 
describes non-variable objects. 
The second and third largest clusters include only 7.7\% and 5.6\% of the 
data, respectively.

The centre of the dominant cluster is (\Ratio, 
\Con, \Neu, \J, \K, \AoVM) = ($7.45\times10^{-3}$, 
$3.90\times10^{-9}$, $1.70$, $8.29$, $7.52\times10^{-1}$, $8.16$), 
which also becomes stationary when the number of clusters converges after the 
10 iterations. The covariance of the multivariate Gaussian model 
for the largest group is 
used for statistical inference to select candidates that may be 
variable sources and will be explained in \S4. Measuring the ratio 
between the covariance of each variability index for the largest cluster and 
that for the whole data of set A, we find that the ratio of \Neu is 0.71 which 
is highest among the six variability indices. Meanwhile, the ratio of \Con is lowest and 
close to zero, suggesting that this variability index has less powerful than 
others in separating out non-variable objects.

\begin{figure}
\includegraphics[scale=0.4]{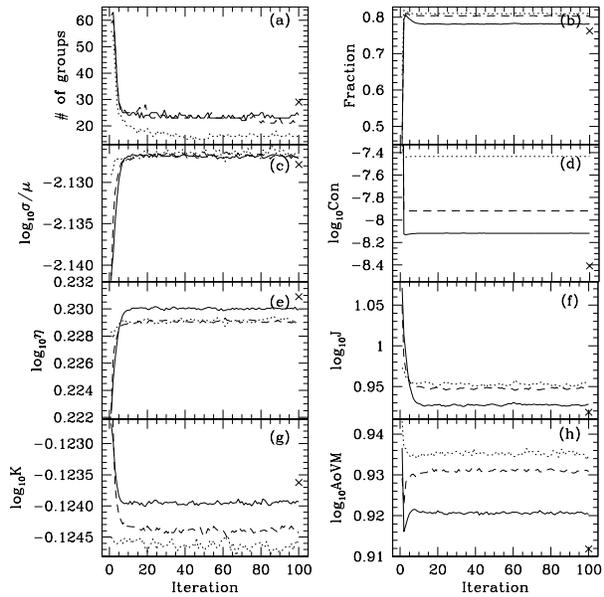}
\caption{The change in cluster properties for randomly selected 
10\%, 30\%, and 50\% samples of set A. In each plot, the cross symbol 
corresponds to the converged values given in Figure \ref{fig:gmm_time_history}. 
(a) The total number 
of identified clusters is smaller than that of the original dataset. 
The subsets are represented as dotted, dashed, and solid lines for 
10\%, 30\%, and 50\%, respectively. (b) 
The largest cluster in the samples has a higher percentage of the data than 
the largest cluster in the original dataset. 
(c - h) For all six variability indices, the result for the 50\% subset 
is most close to what we find for the entire data.}
\label{fig:gmm_subset}
\end{figure}

\subsection{Dependence on the size of data}

\begin{table*}
\begin{minipage}{126mm}
\caption{Changes of the largest group in the 10\%, 30\%, and 50\% samples.}
\label{tab:subset}
\begin{tabular}{@{}ccccc}
\hline
Fraction of data & Included non-variables & 
Recovered non-variables & Included variables & 
Missed non-variables \\
\hline
10\% & 17368 & 17348 (99.9) & 1082 (6.2) & 20 (0.1) \\
30\% & 52043 & 51860 (99.6) & 2856 (5.5) & 183 (0.4) \\
50\% & 86818 & 86070 (99.1) & 2696 (3.1) & 748 (0.9) \\
\hline
\end{tabular}

The numbers in parenthesises show the fraction in percentage 
with respect to the total number of non-variable members (i.e. the second column) that were included in 
the largest group associated with the original dataset and that are also included in the largest 
cluster associated with the subsamples.
\end{minipage}
\end{table*}

\begin{figure}
\includegraphics[scale=0.4]{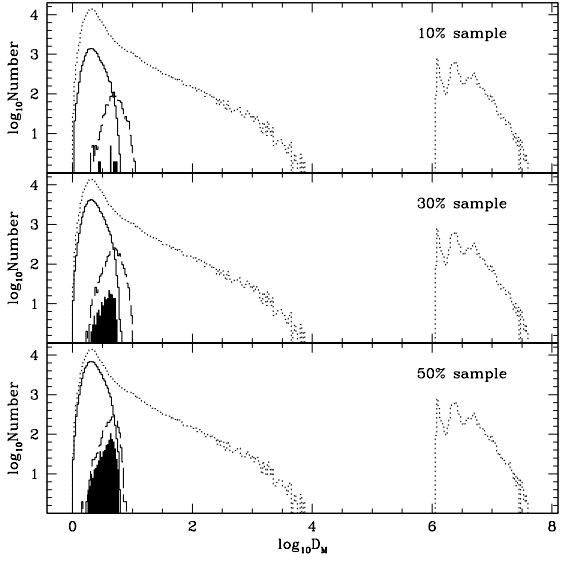}
\caption{
Distribution of the Mahalanobis distances from the largest group with the original data 
for 10\%, 30\%, and 50\% subsamples. In each plot, the distribution for all data points 
(i.e. the original data) is represented by dotted lines, while solid lines are the 
distributions of objects which are included in the largest group associated with both 
the sub-samples and the original data. Objects newly included in the largest group 
associated with the sub-samples ({\it dashed line}) and excluded from that ({\it shaded bar}) are 
mainly from the edge of the original largest group. The distribution represents 
$dN/dlog_{10}D_{M}$ instead of $dN/dD_{M}$.
}
\label{fig:subset_Mahalanobis}
\end{figure}

The dependence of the largest cluster on the size of the 
input data is tested using samples of set A. 
We randomly select 10\%, 30\%, and 50\% of the 
light curves from set A. We compute a GMM 
for these sub-samples using 
the same setup that was used for the main test. Following the basic assumption 
that the largest cluster represents non-variable objects, we identify the largest 
cluster in the three subsets. 
The GMM for the 50\% sample should be closer to the GMM for all of set A 
than the GMMs for 10\% and 30\% samples.

The six variability indices show dependencies 
on the size of the input data. 
In Figure \ref{fig:gmm_subset}, 
the GMM recovers more clusters as the size of dataset increases. Because 
a larger dataset can include more features of data, the GMM 
finds more separable clusters. This result is consistent with our expectation 
for an infinite GMM based on the DP which identifies 
previously unseen structure as the data set with observable features 
increases in size. 
Although each variability index responds differently to the size of data, 
all indices converge more quickly with less data. But the result for 
the 50\% subset shows the better convergence of the indices 
to the original values than other subsets.

However, more iterations do not make it possible to recover more clusters. When we use 
set A, we find 29 clusters with 100 iterations, and 
the number of clusters quickly converges to 29 after 10 iterations. 
But a smaller data produces less clusters more quickly. 
Unsupervised learning techniques naturally handle variation in the size 
of dataset which correspond to variation in the amount of available information. 
Therefore, the maximum number of recovered clusters is not 
dependent on how many times the clustering procedure iterates.

We test the stability of the largest group derived with the original data 
by checking the membership of the largest groups with 10\%, 30\%, and 50\% samples. 
For example, 10\% subsamples have 17368 data points which were included 
in the largest group as shown in 
Table \ref{tab:subset}. Among those data points, 99.9\% of them 
are recovered in the largest group with 10\% subsamples, while 1082 objects 
are newly included in the largest group. However, 20 objects are now enclosed in 
minor groups with 10\% subsamples. In both 30\% and 50\% subsamples, 
the members of the original largest group are well recovered with higher than 99\% rate.
The clustering result is mainly affected by new objects which were not included in the largest group 
associated with the original data, but are included into the largest group 
associated with the sub-samples. 
These data points are mainly from the edge of the largest group in the original data 
as shown in Figure \ref{fig:subset_Mahalanobis}. The 
definition of the Mahalanobis distance $D_{M}$ is
\begin{equation}
D_{M} = \sqrt{(\bf{x} - \bf{\mu}_{0})^{T}\Sigma_{0}^{-1}(\bf{x} - \bf{\mu}_{0})},
\label{eq:D_M}
\end{equation}
where the centre $\bf{\mu}_{0}$ and covariance matrix $\Sigma_{0}$ 
of the largest cluster with the original data are used with the position of an 
individual object $\bf{x}$ in six-dimensional space. Simply, $D_{M}$ corresponds 
to the exponent of the multivariate Gaussian distribution (see Equation \ref{eq:gaussian}). 
Therefore, a high value of $D_{M}$ 
represents a distant object from the centre of the largest group. 
Figure \ref{fig:subset_Mahalanobis} shows that contamination related to the largest group 
is dominated by objects around the edge of the original largest group.

\subsection{Dependence on the noise in data}

\begin{figure}
\includegraphics[scale=0.4]{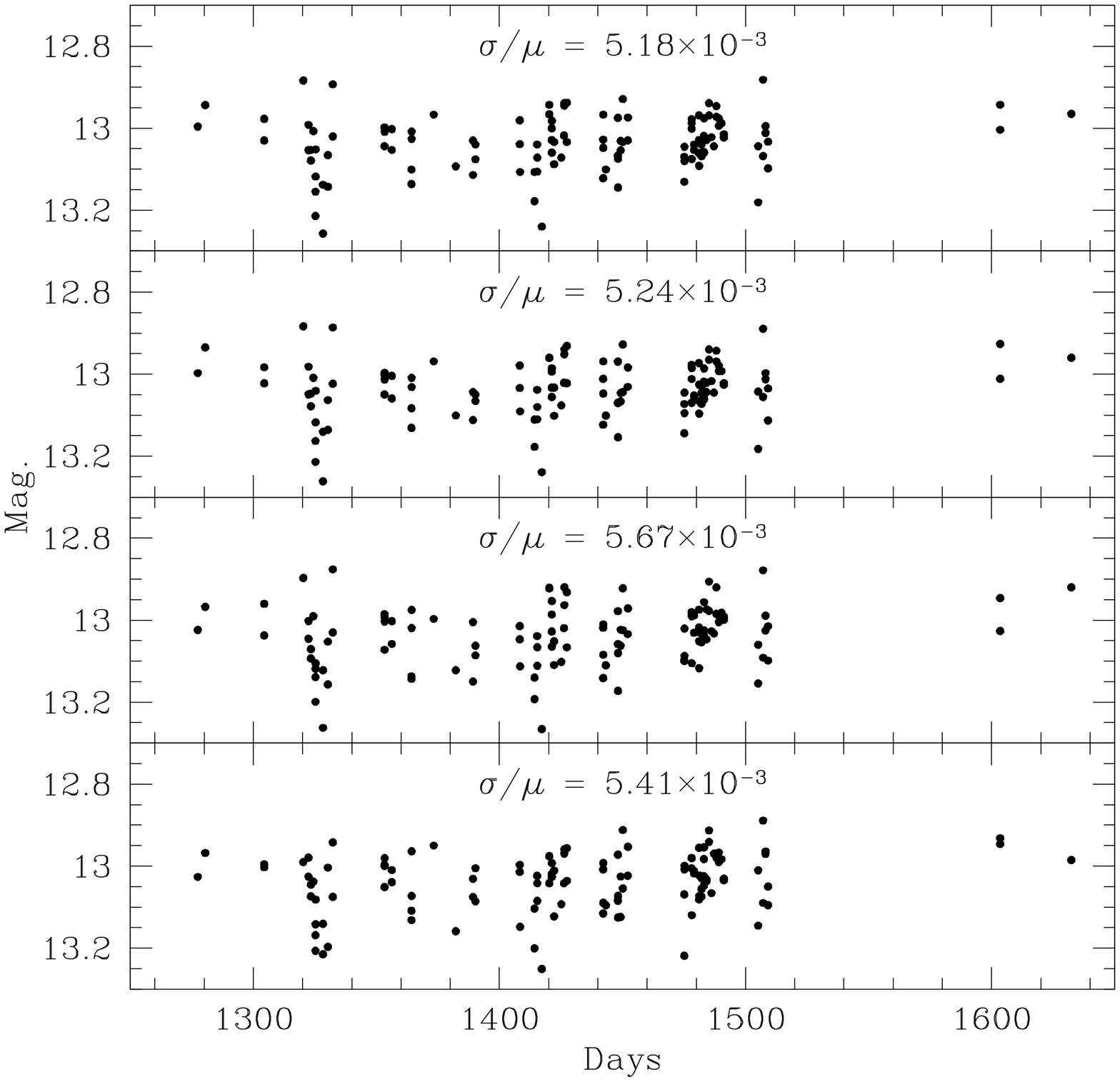}
\caption{Example light curves with increased dispersions. From top to bottom, each 
panel shows the raw light curve and light curves with 10\%, 30\%, and 50\% increased 
dispersions, respectively. The dispersion of the raw light curve is increased by adding 
random values that sampled from a Normal distribution to the existing raw data points.
}
\label{fig:noise_lc}
\end{figure}

\begin{figure}
\includegraphics[scale=0.4]{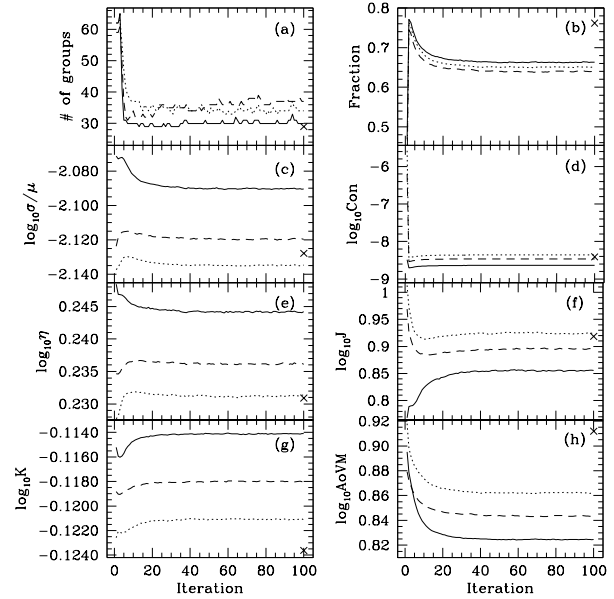}
\caption{The change in cluster properties for set A with 
10\%, 30\%, and 50\% increased dispersions. 
In each plot, the cross symbol 
corresponds to the converged values given in Figure \ref{fig:gmm_time_history}, and 
dotted, dashed, and solid lines represent 10\%, 30\%, and 50\% increased dispersions, 
respectively. 
(a) The total number 
of identified clusters is higher than that of the original dataset. 
(b) The fraction of data in the largest cluster decreases substantially compared with 
the largest cluster associated with the original data. 
(c - h) Six variability indices have different sensitivities 
to the change in magnitude dispersion, implying that 
variability indexes are important to clustering.
}
\label{fig:gmm_noise}
\end{figure}

\begin{table*}
\begin{minipage}{126mm}
\caption{Changes of the largest group in the samples with 10\%, 30\%, and 50\% increased dispersions.}
\label{tab:noise}
\begin{tabular}{@{}cccc}
\hline
Increased dispersions & Recovered non-variables & New non-variables & 
Excluded non-variables \\
\hline
10\% & 146170 (84.4) & 1808 (1.0) & 27017 (15.6) \\
30\% & 141332 (81.6) & 3724 (2.2) & 31855 (18.4) \\
50\% & 143689 (83.0) & 6998 (4.0) & 29498 (17.0) \\
\hline
\end{tabular}

The numbers in parenthesises show the fraction in percentage 
with respect to the total number of members that were included in 
the largest group associated with the original data. 
\end{minipage}
\end{table*}

In order to test the effects of noise on the clustering results, we modify the original data set A 
by adding extra dispersions to the raw light curves. If the magnitude distribution in 
the raw light curves is simply described by the Normal distribution $N(\mu, \sigma^{2})$ with 
mean $\mu$ and dispersion $\sigma^{2}$, we can increase the dispersion of the 
light curve by adding the random number from the Normal distribution $N(0, \sigma_{add}^{2})$ 
to the raw light curve, because the sum of two Normal distribution variables also follow 
Normal distribution:
\begin{equation}
U = X + Y \sim N(\mu_{X} + \mu_{Y}, \sigma_{X}^{2} + \sigma_{Y}^{2}),
\label{eq:normal_dist_add}
\end{equation}
where $X \sim N(\mu_{X}, \sigma_{X}^{2})$ and $Y \sim N(\mu_{Y}, \sigma_{Y}^{2})$. Here, we do not change 
the time sequence of the raw light curve, and use the dispersion of the raw light curve to 
generate the added term with the three cases of $\sigma_{add}^{2} = 0.1\sigma^{2}, 
0.3\sigma^{2}$, and $0.5\sigma^{2}$. These values correspond to 10\%, 30\%, and 50\% 
increases in dispersions, 
respectively. Figure \ref{fig:noise_lc} shows one example light curve which is 
a member of the largest group associated with the original data.

We warn that our approach to degrade the data can be quite different from realistic cases. 
First, there is no guaranty of assuming a Normal distribution for the raw light curves. Second, 
even when raw light curves follow a Normal distribution, the rule given in Equation 
\ref{eq:normal_dist_add} is not well implemented when light curves have a small number of 
data points. Third, if the raw light curve has intrinsic variability which might result in a 
large dispersion, using the dispersion from the raw light curve in Equation \ref{eq:normal_dist_add} 
can cause systematically biased effects on the light curves of truly variable objects. 
Because of these reasons, the increase in dispersion can deviate from 
the expected change in Equation \ref{eq:normal_dist_add} as shown in Figure \ref{fig:noise_lc}. 
Our simulation also fails to reproduce red noise if the data have. 
Even though this test might not be realistic, unsupervised learning 
intrinsically lacks a way to study noise effects without providing completely artificial data.

\begin{figure}
\includegraphics[scale=0.4]{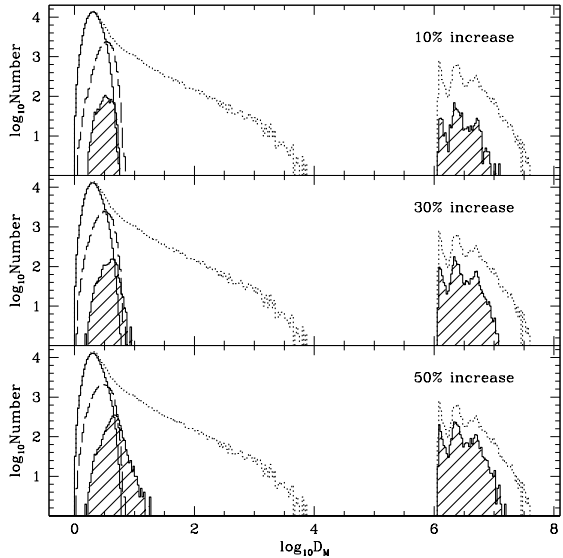}
\caption{
Distribution of the Mahalanobis distances from the original largest group for the samples with 
10\%, 30\%, and 50\% increased dispersions. The distribution for the original data is described 
by dotted lines. For the members of the original largest group, parts of them 
are still included in the largest group with the noise data ({\it solid line}; the second column in 
Table \ref{tab:noise}) even after they were altered by added dispersions. 
But as 
the dispersion increases, more objects ({\it shaded histogram}; the third column in 
Table \ref{tab:noise}) are newly included in the largest group with the noise data, 
while some members of the largest group with 
the original data ({\it dashed line}; the fourth column in 
Table \ref{tab:noise}) are now excluded from the new largest group.
}
\label{fig:noise_Mahalanobis1}
\end{figure}

Figure \ref{fig:gmm_noise} summarises the effects of noise on clustering and the largest group. 
The increase in noise enhances dispersions among clusters that were found with the original data, and 
results in the recovery of more clusters because the largest group 
is populated by fewer objects. 
Importantly, variability indices associated with the centre 
of the largest group responds to the effects of noise in different ways. 
Therefore, the change of the cluster centre for the largest group 
is not a simple function of the dispersion change although the data with the low noise 
generally converges to the results for the original data. 

\begin{figure}
\includegraphics[scale=0.4]{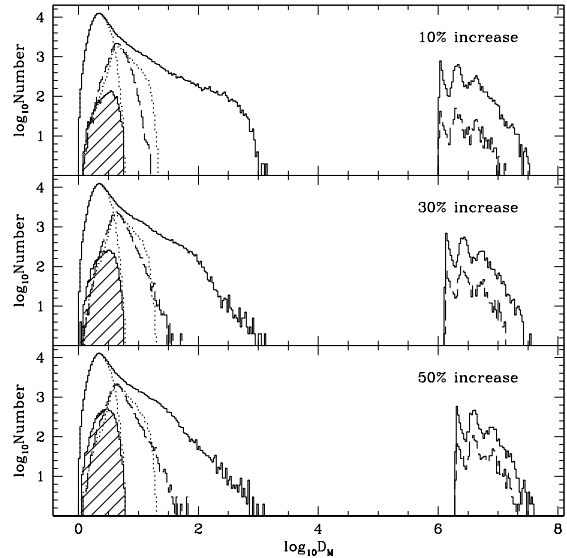}
\caption{
Distribution of the Mahalanobis distances from the new largest group in the samples with the 
increased dispersions. The solid line shows the distribution of all objects with respect to 
the new largest group. The dashed line and shaded histogram represent the same objects as 
explained in Figure \ref{fig:noise_Mahalanobis1}. The top two largest clusters are shown by 
dotted lines for each case.
}
\label{fig:noise_Mahalanobis2}
\end{figure}

We also trace which objects are included in the newly found largest cluster. 
The added dispersion naturally boosts mixing between the original largest group and other 
minor groups. As presented in Table \ref{tab:noise}, about 16\% - 18\% of objects that were 
included in the largest group associated with the original data are found in minor groups with the 
increased dispersions. Meanwhile, the addition of new objects to the largest group is a small 
percentage. 
Figures \ref{fig:noise_Mahalanobis1} and 
\ref{fig:noise_Mahalanobis2} show that the increased dispersions induce 
the objects around the edge of the original largest cluster to move from the largest cluster 
in the new clustering. 
Figure \ref{fig:noise_Mahalanobis2} demonstrates that this effect mainly 
results in grouping objects into the second largest cluster in the new data.

\subsection{Dependence on the source of data: results of set B}

\begin{figure}
\includegraphics[scale=0.4]{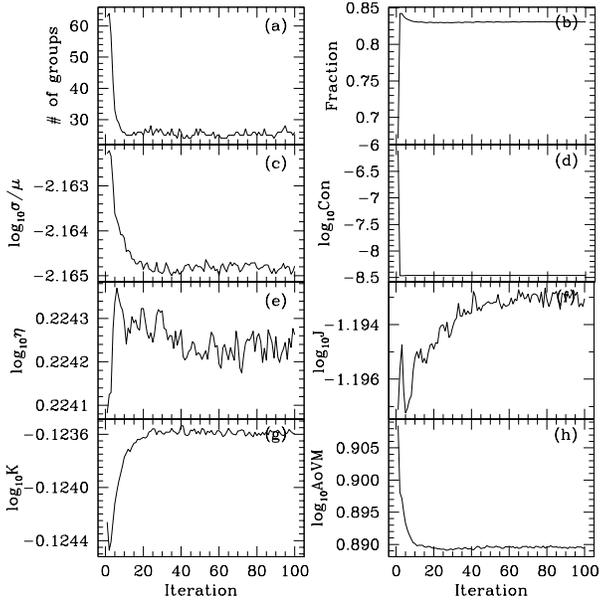}
\caption{The change in cluster properties for set B. 
The plotted fields are the same as those in Figure \ref{fig:gmm_time_history}. 
We find some difference in the largest cluster between sets A 
and B as we expect in data-driven machine learning. 
In particular, $J$ shows the most significant difference.
}
\label{fig:gmm_time_test_history}
\end{figure}

We test the dependence of the GMM on the properties 
of a particular dataset by applying our method to 
set B. As described in \S2.1, set B has different properties of data. 
The largest cluster of set B is populated by 83.1\% of data, 
while we find that the largest cluster of set A 
is populated by 76.2\% of the data 
(see Figure \ref{fig:gmm_time_test_history}). 
The number of recovered clusters is 26 which is smaller than that of set A. 
Even though fewer clusters are identified in set B, 
the largest cluster in set B describes more of the data. 

Figure \ref{fig:gmm_time_test_history} shows how each variability index changes based 
on the input data. In this test, $J$ is a signature of a large change that depends 
on the properties of the data. We note that $K$ or $AoVM$ is a variability index 
that shows the largest change for the noisy data (see Figure \ref{fig:gmm_noise}). 
The centre of the largest cluster in set B is 
(\Ratio, \Con, \Neu, \J, \K, \AoVM) = 
($6.84\times10^{-3}$, $3.45\times10^{-9}$, 1.68, $6.41\times10^{-2}$, 
$7.52\times10^{-1}$, 7.75). This result implies that our method has to be 
applied to a single dataset that shares common properties. 
This requirement is often necessary for 
data-oriented machine learning methods. 
Compared to the test associated with increasing the dispersion of light curves, the 
experiment with set B is more realistic in proving the data-dependence of unsupervised 
machine learning algorithms.

\section{Separation of variable objects}

After we identify the largest cluster as the aggregation of non-variable objects, the next 
question is how to separate out candidates of variable objects. Naively, we can accept 
the clustering results of the GMM as a guide line for the separation. 
But simply depending on clustering results is not satisfactory for two reasons. 
First, any systematic 
spurious patterns can be clustered as the second or third largest cluster, 
as seen in Figure \ref{fig:noise_Mahalanobis2}. Second, 
some clusters can be close to the largest cluster in six-dimensional space, 
implying that the separation of other clusters from the largest cluster might not be meaningful. 
Therefore, if the clustering result is used to define candidates of variable objects, 
then various classical methods for multivariate analysis can be applied \citep{krzanowski88} 
in addition to the cluster membership from the GMM with the DP. 
We suggest two simple ways to use the results from our GMM method with the 
clustering membership. 

\subsection{Inference from Mahalanobis distances}

The first approach uses the 
Mahalanobis distance 
to gauge how far an object is from the largest cluster. For 
our application, the Mahalanobis distance is more useful 
than a multidimensional norm because it includes 
the effects from the dispersion of the data \citep{bishop06}.

\begin{figure}
\includegraphics[scale=0.4]{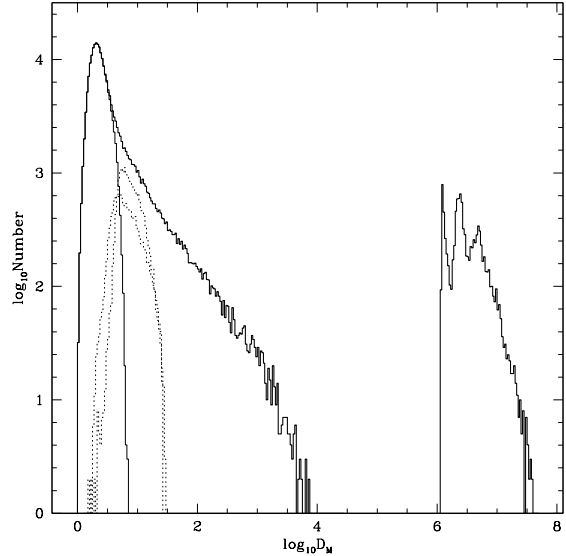}
\caption{Mahalanobis distance of all objects in set A. The distribution of all 
objects ({\it thick solid line}) has a peak around $D_{M} \sim 2$ which 
corresponds to the distribution of only the largest cluster ({\it thin solid line}). 
The second and third largest clusters ({\it dotted lines}) are distributed closely to 
the largest cluster, even though they are identified separately by the GMM with the DP.
}
\label{fig:Mahalanobis_dist}
\end{figure}

The distribution of $D_{M}$ shown in Figure \ref{fig:Mahalanobis_dist} indicates that 
a cut based on $D_{M}$ can be used to identify variable objects. 
This distribution has a concentration of objects around $D_{M} \sim 2$. The 
position of this peak matches the mode value of the Beta distribution which 
is expected for the distribution of $D_{M}$ \citep{ververidis08}. 
This distance also represents the typical distance of objects that are included in 
the cluster of non-variable objects (i.e. the largest cluster). 
Furthermore, this distribution confirms 
that the second and third largest clusters may not represent real variable objects 
because the members of the clusters are close to the largest cluster.

Even though $D_{M}$ is inexpensive to compute, it does not give direct statistical 
inference nor provide a 
statistical confidence limit on our belief that an object is variable. $D_{M}$ 
is simply an exponent of the multivariate Gaussian distribution (see Equation \ref{eq:D_M}). 
One has to find 
an empirical cut for $D_{M}$ that separates variable and non-variable objects.

\subsection{Confidence bounds}

The way to extract a direct statistical inference is 
to derive confidence bounds for 
non-variable objects with $D_{M}$. With the identified 
centre $\bf{\mu_{0}}$ and covariance matrix $\Sigma_{0}$ 
of the largest cluster, we define a confidence bound of 100$b$\% (0 $< b <$ 1) which 
encompasses 100$b$\% of non-variable objects \citep[see]
[for an example]{chen06}. The confidence bound is described as a likelihood threshold 
$h$ that is associated with the probability $b$:
\begin{equation}
\int_{\bf{x}:p(\bf{x})>h} p(\bf{x} \vert \bf{\mu_{0}}, \Sigma_{0}) d\bf{x} = b,
\label{eq:integration}
\end{equation}
where $p(\bf{x})$ is a multivariate Gaussian distribution 
defined by $\bf{\mu_{0}}$ and 
$\Sigma_{0}$. From this integration, we can 
estimate a confidence limit that corresponds to a specific $D_{M}$ for $h$. 
But despite its statistical robustness, 
this integration is practically difficult and expensive to compute 
because it cannot be calculated analytically.

We use a Monte Carlo method to find the confidence limit 
in Equation \ref{eq:integration}. 
An approximate cut is simply the value of $D_{M}$ that 
includes 100$b$ \% of the data in the largest cluster, when 
sorting $D_{M}$ in an ascending order, i.e. 
a descending order of $p$. 
However, a more precise estimate is made possible by generating 
multiple samples of the data that populate 
the largest cluster and finding a limit 
for $D_{M}$ in each sample \citep{chen06}. 
In Figure \ref{fig:Mahalanobis_dist}, 
we can guess that $D_{M} = 4.67$ for set A 
where an approximate cut of 99\% is assumed. 
But we find 
$D_{M} = 4.68$, when using the Monte Carlo method 
by sampling 50 times with 2000 samples for each sampling. 
Because the largest group includes a large number of data points, the simple approximation is 
close to the estimate given by the Monte Carlo method. 
When we choose a 90\% cut in $D_{M}$ to define variable source candidates, 
the total number of candidates is 50,394 for set A and corresponds to 
about 22\% of the light curves. But we find that 
about 9\% and 29\% of objects in the second and third largest group are 
within the 99\% cut of the largest group (i.e. $D_{M} = 4.68$).

\subsection{Examples of light curves for each cluster}

Our method provides two pieces of information to help people select variable source candidates. 
First, objects are chosen as candidates when they are not included 
in the largest cluster. This idea corresponds to a classical method of outlier detection which 
uses clustering. Second, we use the cut $D_{M}$ in addition to the result of 
clustering. As shown in Figure \ref{fig:Mahalanobis_dist}, the second and third largest 
clusters are overlapped with the largest cluster in the six-dimensional space. This second 
approach is relevant to the distance-based outlier detection \citep{cateni08}. 
The most useful approach is to employ both the clustering results and the statistical cut in $D_{M}$. 
If we conclude that the second and third largest clusters are explained by 
a systematic bias in the data, then we can exclude the second and third largest clusters when 
selecting variable source candidates. Furthermore, $D_{M}$ can be used to assign 
a priority to the candidates.

\begin{figure}
\includegraphics[scale=0.4]{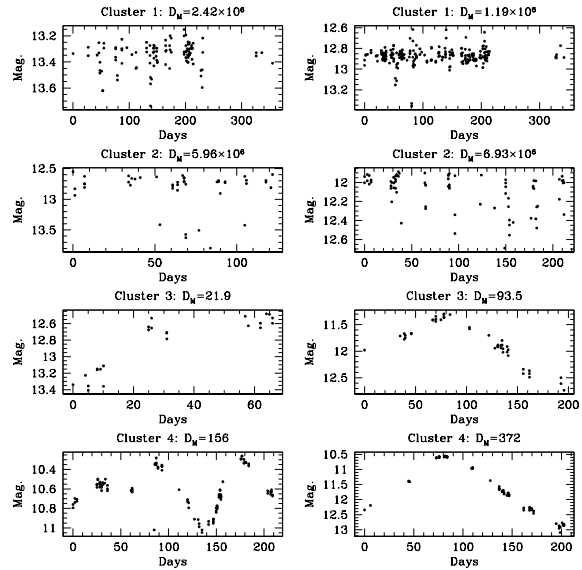}
\caption{An example of light curves for clusters 1, 2, 3, and 4. The second light curve  
for cluster 3 with $D_{M}= 93.5$ is matched to a known variable star 
SV* BV 1711 \citep{strohmeier75}, while the second light curve for cluster 4 is 
IRAS 19225-0740 \citep{helou88}.
}
\label{fig:example_cluster1_4}
\end{figure}

\begin{figure*}
\includegraphics[scale=0.4]{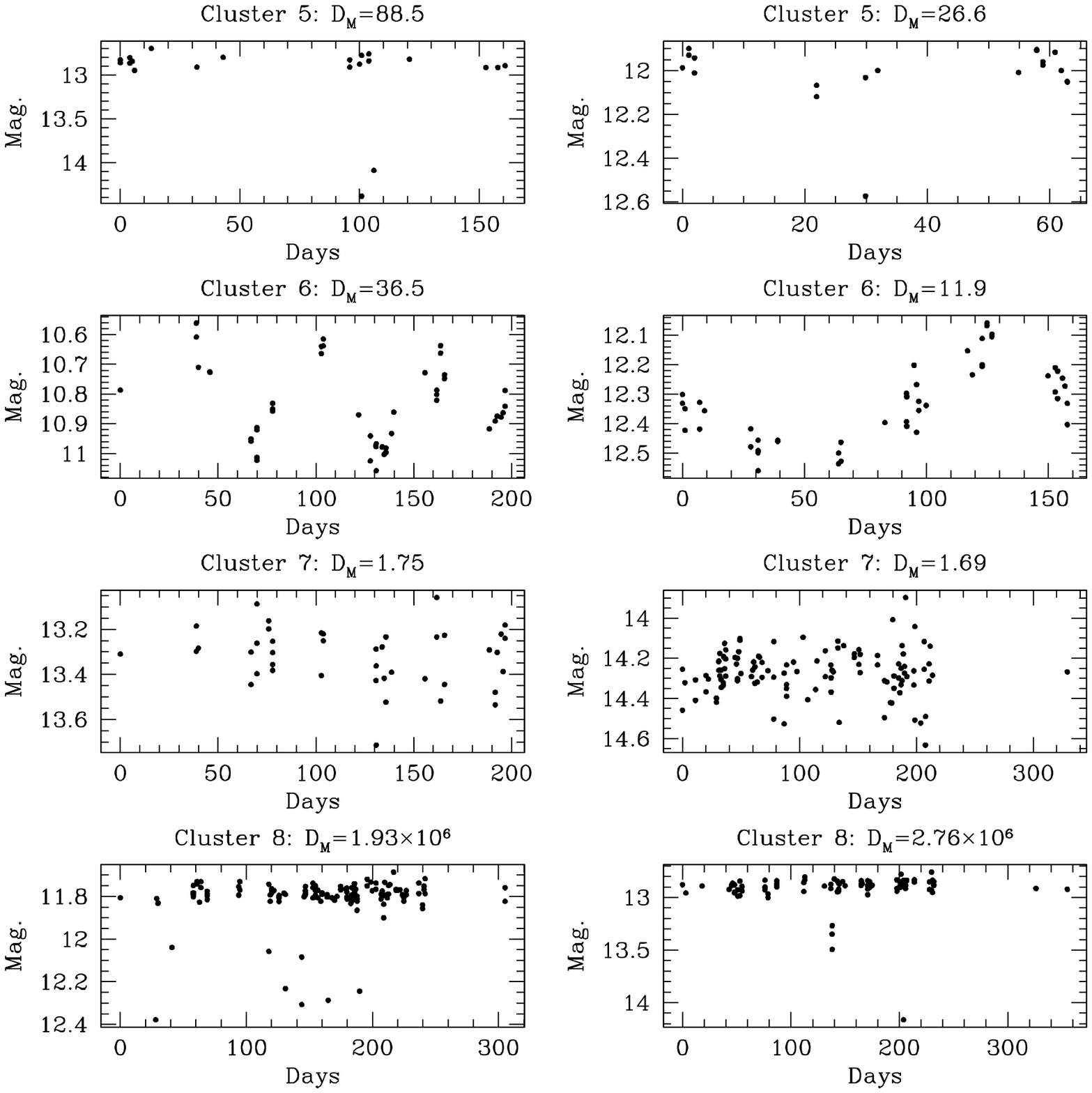}
\includegraphics[scale=0.4]{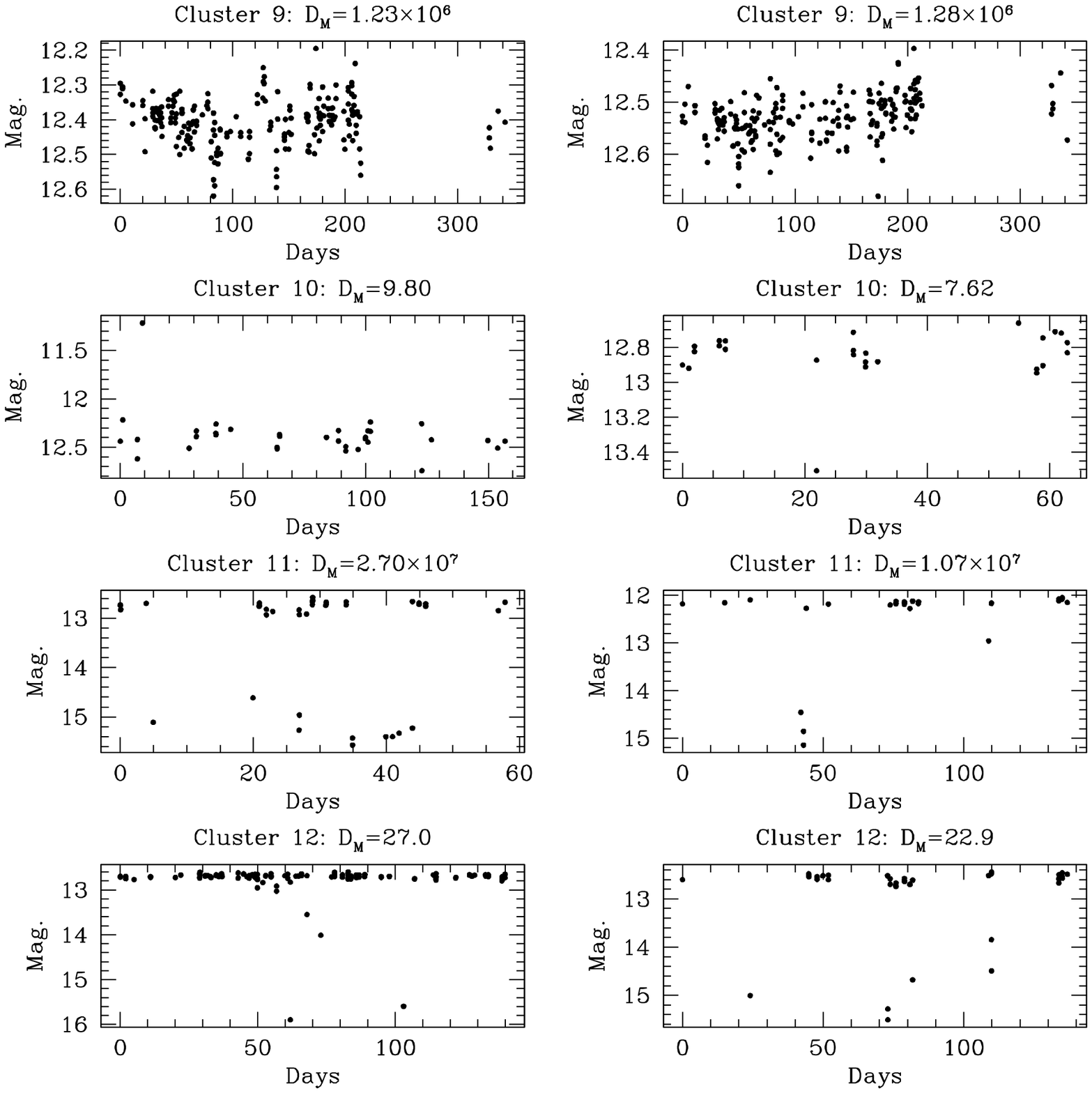}
\caption{Example light curves for clusters 5 - 12. None of these examples are 
known variable stars. Cluster 7 is the largest cluster which represents 
non-variable objects. Both light curves of cluster 6 show a recognisable 
change in brightness even with poor sampling of the light curves. Cluster 10 
is the second largest cluster that has some objects within 
the 99\% cut of $D_{M} \sim 4.7$
}
\label{fig:example_cluster5_12}
\end{figure*}

\begin{figure*}
\includegraphics[scale=0.4]{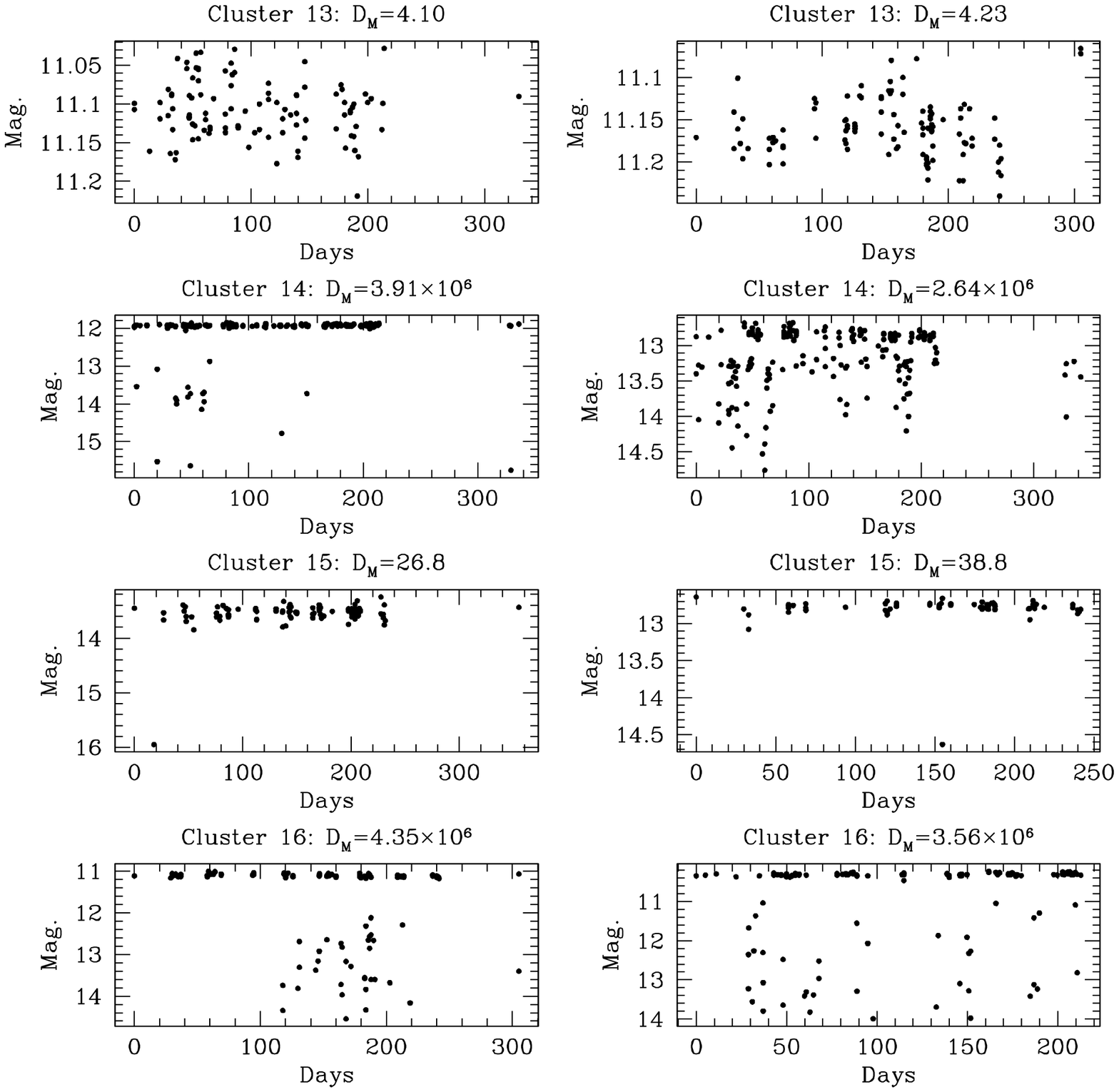}
\includegraphics[scale=0.4]{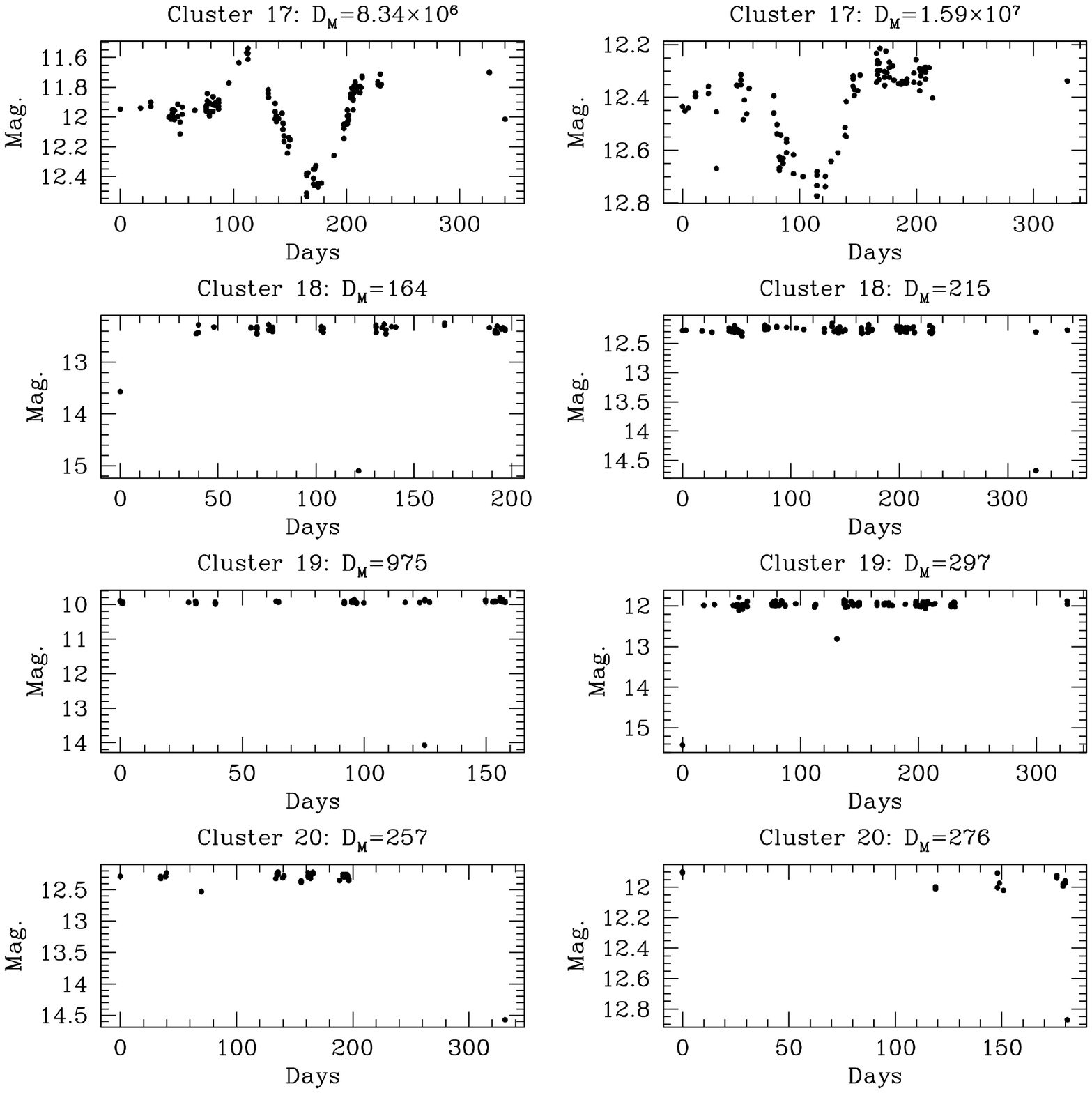}
\caption{Example light curves for clusters 13 - 20. No known variable sources 
were found for these example objects within the 6\farcs0 search radius using 
the SIMBAD. But both examples of cluster 17 show long-period variability. 
Additionally, the examples of cluster 16 might be eclipsing binaries. Cluster 
13 is the third largest cluster, and includes about 6\% data of set A. 
}
\label{fig:example_cluster13_20}
\end{figure*}

\begin{figure*}
\includegraphics[scale=0.4]{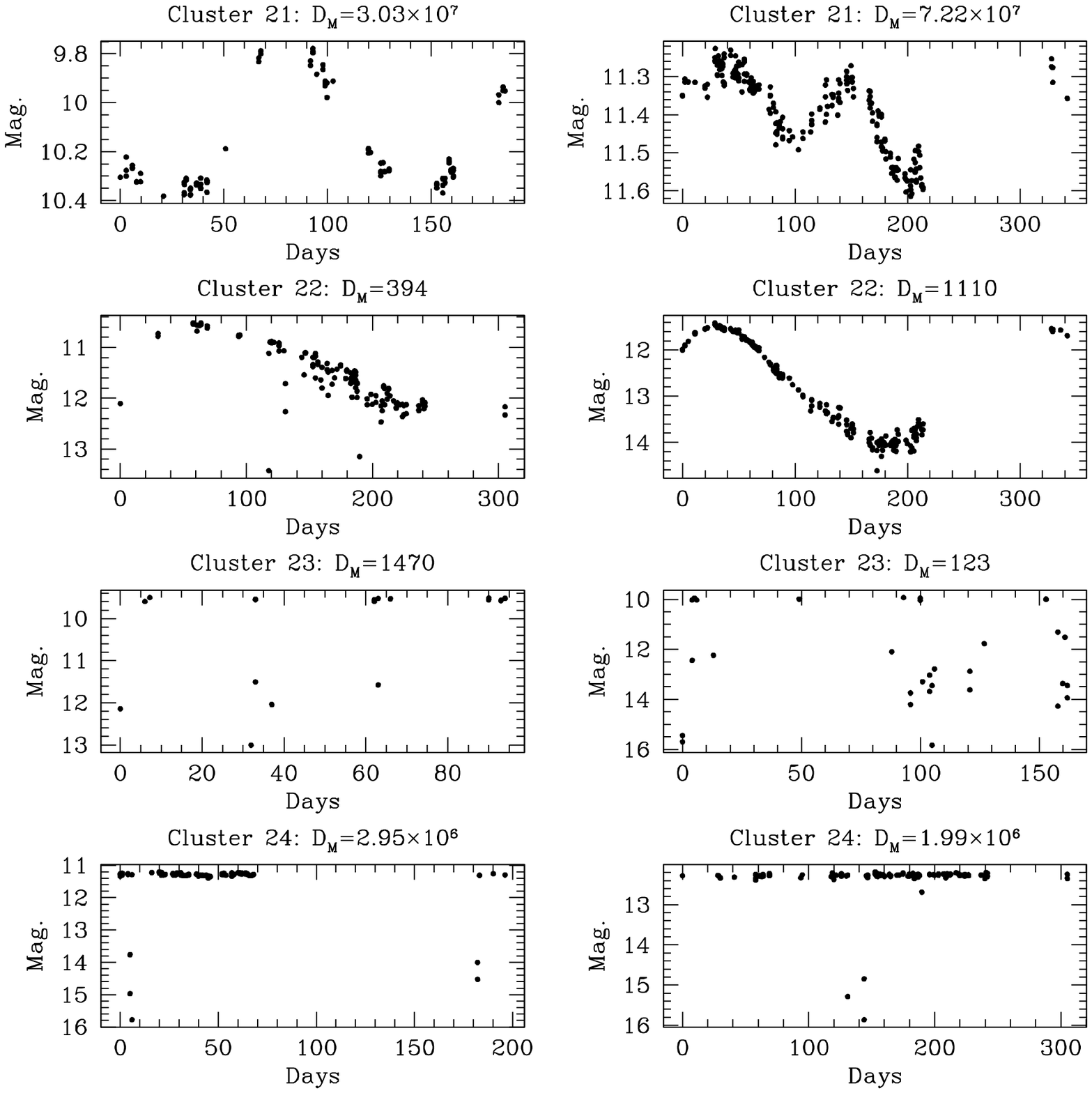}
\includegraphics[scale=0.4]{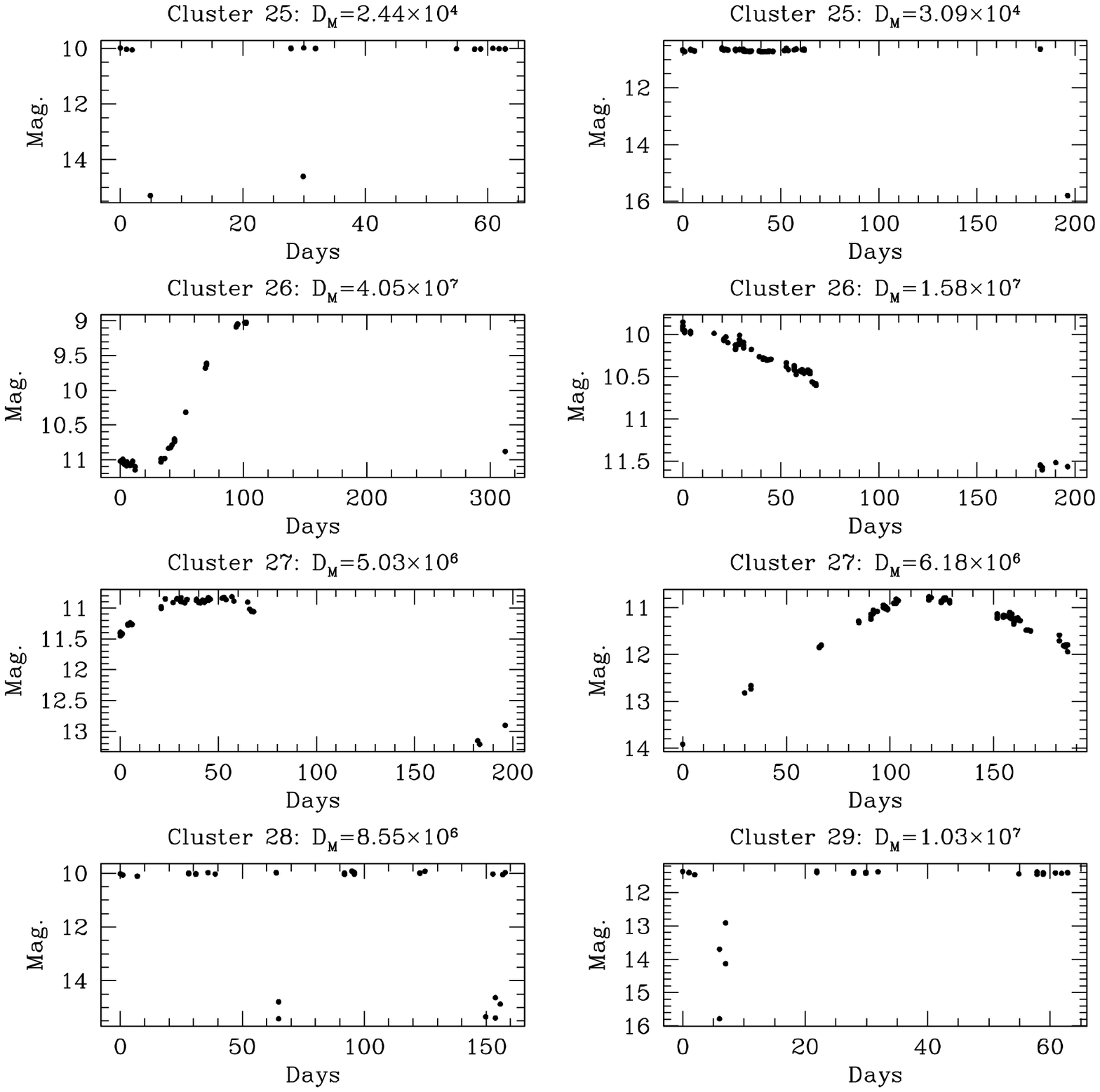}
\caption{Example light curves for clusters 21 - 29. Except for clusters 28 and 
29 which have only one member, two examples are presented for each cluster. The first 
example of cluster 21 spatially corresponds to IRAS 20302+1938 \citep{helou88}. 
The second example with $D_{M}=7.22\times10^{7}$ is the infrared carbon star 
IRAS 18364+1757 \citep{helou88,guglielmo93}. While the first example of cluster 
22 is the known variable star V* V2328 Cyg \citep{dahlmark01}, the second example is 
not a known variable object even though its light curve shows a clear sign of 
variability. For cluster 26, the first example is a Mira-type variable star 
V* Z Del \citep{templeton05}.
}
\label{fig:example_cluster21_29}
\end{figure*}

Figure \ref{fig:example_cluster1_4} presents an example of light curves for clusters 1, 2, 
3, and 4 in set A. Here, we randomly select two light curves for each cluster. 
Cluster 1 is the fourth largest cluster in the GMM for set A. In order 
to check for known variable sources in our samples, we spatially match our samples 
to the SIMBAD database \citep{simbad} using a 6\farcs0 search radius. For 
cluster 3, the second object is the known variable star SV* BV 1711 \citep{strohmeier75}. 
The second object for cluster 4 is the known infrared source IRAS 19225-0740 
\citep{helou88} which may be a long-period late-type variable star.

For the rest of the identified clusters in set A, we also randomly extract two example objects. 
These light curves are presented in Figures \ref{fig:example_cluster5_12}, 
\ref{fig:example_cluster13_20}, and \ref{fig:example_cluster21_29}. Only a small number of 
objects among the examples are known variable stars or infrared sources that might be 
long-period variable stars. The clusters 10 and 13, corresponding to the second and third 
largest cluster respectively, show similarities in their light curves to those of the largest cluster 
(i.e. the cluster 7). 
Because of poor sampling for short-period variable sources in the NSVS, it is not likely 
for us to recognise periodic short variability in the example light curves.

\section{Discussion and Conclusion}

We presented a new framework for discovering variable objects in massive time-series data 
with variability indices which have been commonly used in astronomy. 
Our method is fully non-parametric and depends on only one assumption: the 
largest cluster represents a group of non-variable objects. The infinite GMM with the DP 
derives a mixture of multivariate Gaussian distributions from the given data consisting 
of six variability indices. With these results, 
we use the clustering results and Mahalanobis distances from the largest 
cluster to select variable object candidates.

Our application of the infinite GMM with the DP for clustering may be useful 
for measuring how 
efficiently we recover variable objects depending on various factors. Before 
designing the observation strategy to acquire time-series data, simulated data can be 
applied to our method. This test will help people understand what kind of variability 
is missed in the data given by a specific observation strategy and environment.

Pre-processing the time-series data, i.e. feature extraction and selection, 
can affect the results of the infinite GMM with the DP. 
This effect has been seen in other unsupervised clustering algorithms, too 
\citep{jain99}. In this paper, we only use six variability indices which 
are mainly developed for astronomical time-series data. Unlike time-series data 
in other fields, astronomical time-series data are irregularly sampled and less 
homogeneous. This difference makes pre-processing of our data with a common method 
such as principal component analysis difficult. Moreover, finding the best features 
for unsupervised clustering is a trial-and-error problem \citep{jain99}. In the 
framework of the GMM with the DP, the importance of each variability index is 
reflected in each Gaussian component's covariance matrix which 
also describes the compactness of the found clusters. Therefore, 
the combination of the best features will be dependent of 
the input data, while making this study be a trial-and-error problem \citep{dy04}. 
In the next paper of this series, we will investigate the usefulness of a variety of 
variability indices for the infinite GMM with the DP.

Finally, simply selecting variable star candidates with the unsupervised learning method 
is not useful without analysing what kind of objects are selected as candidates. 
Our approach also uses unlabelled data which 
we do not know any physical properties about. It is necessary to figure out properties of 
the selected candidates in the further analysis of variable time-series data.

We showed that our method is a fully data-driven approach such that the method 
itself finds its best separation of variable objects for the given data. This 
property makes our idea easily applicable to future projects such as Pan-STARRS \citep{kaiser04}, 
GAIA \citep{eyer00}, and LSST \citep{walker03} as well as archives of the past surveys such as 
MACHO \citep{cook95}. In another paper, we will provide a full list of 
variable object candidates from the NSVS for each observation field.

\section*{Acknowledgements}

We are grateful to David Blei for useful discussions and 
Przemek Wozniak for helping us to extract sample data from the NSVS 
database. We also thank the referee for useful comments which help us 
improve the paper substantially. 
M.-S. is supported by the Charlotte Elizabeth Procter Fellowship of Princeton 
University. 
M.-S. is also partly 
supported by the Korean Science and Engineering Foundation Grant KOSEF-2005-215-C00056 
which is funded by the Korean government (MOST). 
M.S. acknowledges support from the DOE CSGF Program which is provided 
under grant DE-FG02-97ER25308.

\newcommand{\bra}[1]{\ensuremath{\left\langle #1 \right|}}
\newcommand{\ket}[1]{\ensuremath{\left|#1\right\rangle}}
\newcommand{\braket}[2]{\ensuremath{\left\langle#1 |  #2\right\rangle}}
\newcommand{\tensop}[3]{\ensuremath{#1^#2_#3}}
\newcommand{\partdif}[2]{\ensuremath{ \frac{\partial #1}{\partial #2}}}

\appendix
\section{Bayesian Non-parametric Clustering}

Bayesian nonparametric clustering algorithms based on the 
Dirichlet Process are a powerful way to model and manipulate 
data in statistics, machine learning, and signal processing. 
In this type of analysis, Bayesian refers to the manner in 
which one estimates the likelihood of an event given 
information in the data set about all known events and 
nonparametric refers to the manner in which a set of 
events can be modelled such that the structure of the model 
is determined only by the data set. Since Bayesian nonparametric 
techniques are not based on prior assumptions about the 
structure, number of mixture components, or location of components 
in a data set, one employs the Dirichlet Process to assign a prior 
probability (i.e., the unconditional probability of an event before 
relevant information is considered) to a data point $x_{n}$ such 
that the stochastic generative process draws from a distribution 
of distributions (in our case, a mixture of multivariate Gaussian 
distributions) \citep{ferguson73, antoniak74, jordan05}. 

The Dirichlet Process mixtures are also referred to as infinite 
mixtures because although data may exhibit a finite number of components, 
new data can exhibit previously unseen structure 
\citep{neal00, blei04}. Therefore, these models adjust 
their complexity according to the complexity of the data and mitigate 
under-fitting the data \citep{teh07}. In this unsupervised algorithm, 
no data points were discarded as background.

To understand the Dirichlet Process and our Bayesian 
nonparametric clustering algorithm, we explain the 
following ideas from probability theory. Let $\eta$ be a 
probability space, $G_{0}$ be a distribution over $\eta$, 
and $\alpha$ be a positive real number (in our case, $\alpha = 1$). 
Therefore, a random distribution $G$ over $\eta$ is said to be 
Dirichlet Process distributed:
\begin{equation}
G \sim \textrm{DP}(\alpha, G_{0}),
\end{equation}
if and only if for all natural numbers $j$ and any finite partition 
$(A_{1},\ldots,A_{j})$ of $\eta$, the random vector 
$(G(A_{1}),\ldots,G(A_{j}))$ is distributed as a finite-dimensional 
Dirichlet distribution:
\begin{equation}
(G(A_{1}),\ldots,G(A_{j})) \sim \textrm{Dir}(\alpha G_{0}(A_{1}), \ldots, \alpha G_{0}(A_{j})),
\end{equation}
where $G_{0}$ is the base distribution of $G$ 
(i.e., mean of the Dirichlet Process) and $\alpha$ 
is the concentration parameter (i.e., inverse variance of the 
Dirichlet Process) \citep{blei04,jordan05,teh07}.

Bayesian nonparametric clustering based on the Dirichlet 
process can be applied to $N$-dim data with multiple parameter 
fields $(x_{1},\ldots,x_{N})$ provided that the data is regarded 
as being part of an indefinite exchangeable sequence. One models 
the distribution from which $x$ is drawn as a mixture of distributions 
of the form $F(\eta)$, with the mixing distribution over $\eta$ being $G$, 
which has the Dirichlet Process as a nonparametric prior probability. 
Therefore, the Dirichlet Process mixture model is represented as 
\citep{ferguson73, antoniak74, neal00, blei04, teh07}:
\begin{eqnarray}
G                & \sim & \textrm{DP}(\alpha, G_{0}) , \\
\eta_{m} | G     & \sim & G , \\
x_{n} | \eta_{m} & \sim & F(\eta_{m}).
\end{eqnarray}

Since the parameters $\eta$ are drawn from $G$, the data $x$ clusters 
according to the values of $\eta$. For the cluster model presented in 
this work, $x$ is drawn from $F$, which is assumed to be a mixture of 
multivariate Gaussian distributions. Therefore, 
$\eta_{m} \rightarrow (\mu_{m}, \Sigma_{m})$, where $\mu_{m}$ 
is the mean and $\Sigma_{m}$ is the covariance matrix for the $m^{th}$ mixture component.

In Dirichlet Process mixture modelling, the posterior distribution on the partitions 
(i.e., the conditional probability of the mixture components) is intractable 
to compute. However, Markov chain Monte Carlo methods allow one to approximate 
posteriors by constructing a Markov chain that is easy to implement for 
models based on conjugate prior distributions such as the Gaussian 
distributions used in this paper \citep{neal00, blei04}. 
The most widely used inference method is the Gibbs sampler because of 
its simplicity and good predictive performance. In the Gibbs sampler, 
the Markov chain is obtained by iteratively sampling each variable that 
is conditioned on the data and other previously sampled variables. 
If one integrates out all random variables except $q_{m}$ 
(i.e., mixture component that the $n^{th}$ data point $x_{n}$ is 
associated with), then one arrives at the collapsed Gibbs sampler, which 
iteratively draws each $q_{m}$ from the following expression \citep{blei04}:
\begin{equation}
p(q_{m}=1 | x, q_{-m}, \lambda, \alpha) \propto 
p(x_{n} | x_{-i}, q_{-m}, q_{m}=1, \lambda) p(q_{m}=1 | q_{-m}, \alpha),
\label{eq:draw}
\end{equation}
where $q_{-m}$ denotes all of the previously sampled cluster variables 
except for the $m^{th}$ variable and $\lambda$ is a hyper-parameter that 
is used to define the base distribution $G_{0}$. The first term on the 
right-hand side of Equation \ref{eq:draw} is a combination of normalising 
constants that comes from considering Dirichlet Process mixtures for which 
data is drawn from an exponential family (e.g., Gaussian distribution). 
The second term on the right-hand side is:
\begin{equation}
p(q_{m}=1 | q_{-m}, \alpha) = \left\{ \begin{array}{ll}
           \frac{n_{m}}{N-1+\alpha}  & ~~ \textrm{seen component} \\
           \frac{\alpha}{N-1+\alpha} & ~~ \textrm{unseen component} \end{array} \right.,
\label{eq:crp}
\end{equation}
where $n_{m}$ is the number of members in $q_{m}=1$. 
Equation \ref{eq:crp} comes from the partition 
structure of the Dirichlet Process and is the heart of the 
algorithm's clustering effect such that the more frequently 
an event (i.e., a mixture component) is sampled in the past 
- the more likely the event is to be sampled in the future 
\citep{blei04}. Once the Markov chain has run for a 
sufficiently long duration, samples of $q$ will be samples 
from $p(q | x, \alpha, \lambda)$ and one can construct an 
empirical distribution to approximate the posterior.

The collapsed Gibbs sampler runs for a specified number of 
iterations and also iterates over the number of data items $N$. 
The method proceeds with the following steps \citep{teh07}:
\begin{enumerate}
\item Remove data item $x_{n}$ from component $q_{m}$, where $m$ specifies the cluster to which data item $n$ belongs
\item Delete the active component $q_{m}$ if it has become empty
\item Compute conditional probabilities $(p_{1},\ldots,p_{M})$ with respect to data item $x_{n}$ belonging to each of the $M$ active components $(q_{1},\ldots,q_{M})$
\item Choose new component identity $m$ by sampling from the conditional probabilities
\item If $m=M+1$, then create a new active component
\item Add data item $x_{n}$ into component $q_{m}$
\end{enumerate}

\end{document}